\documentclass{aastex61}

\received{}
\revised{}
\accepted{}
\submitjournal{ApJ}

\shorttitle{Quasi-Keplerian Accretion Disc Around Magnetized Stars}
\shortauthors{Habumugisha et al.}
\begin{document}

\title{The Structure of a Quasi-Keplerian Accretion Disc Around Magnetized Stars}

\correspondingauthor{Isaac Habumugisha}
\email{hisaac08@yaoo.co.uk}

\author[0000-0002-0786-7307]{Isaac Habumugisha}
\affil{Department of Physics, Mbarara University of Science and Technology, Mbarara, Uganda}\affil{Department of Physics, 
Kabale University, Kabale, Uganda}\affil{Department of Physics, 
Islamic University in Uganda, Mbale, Uganda}

\author{Edward Jurua }
\affiliation{Department of Physics, Mbarara University of Science and Technology, Mbarara, Uganda}
\nocollaboration

\author{Solomon B. Tessema}
\affiliation{Astronomy and Astrophysics Research Division, Entoto Observatory and Research Center, Addis Ababa, 
Ethiopia}

\author{Anguma K. Simon}
\affiliation{Department of Physics, Muni University, Arua, Uganda}

% \author{Jeff Lewandowski}
% \affiliation{IOP Senior Publisher for the AAS Journals}
% \affiliation{IOP Publishing, Washington, DC 20005}



\begin{abstract}

In this paper, we present a complete structure of a quasi-Keplerian thin accretion disc with an 
internal dynamo around a magnetized neutron star. We assume a full quasi-Keplerian disc with the azimuthal velocity 
deviating from the Keplerian fashion by a factor 
$\xi~(0<\xi<2)$. In our approach, we vertically integrate the radial component of the momentum equation to obtain the 
radial pressure gradient equation for a thin quasi-Keplerian accretion disc. Our results show that at large radial 
distance the accretion disc behaves in a Keplerian fashion. But, close 
to the neutron star, pressure gradient force (PGF) largely modifies the disc structure resulting into sudden dynamical 
changes in the accretion disc. 
The corotation radius is shifted inwards (outwards) for $\xi>1$ (for $\xi<1$) and the position of the 
inner edge with respect to the new corotation radius is also relocated accordingly as compared to the Keplerian model. 
The resulting PGF torque couples with viscous torque (when $\xi<1$) to provide a spin-down torque and a spin-up torque 
(when $\xi>1$) while in the advective state. 
 Therefore, neglecting the PGF is a big omission as has been the case in 
previous models.
This result has the potential of explaining the observable dynamic consequences of accretion discs around magnetized 
neutron stars.

\end{abstract}

\keywords{Accretion discs --
                Pressure gradient --
                Magnetic dynamo -- 
                quasi-Keplerian}



\section{Introduction} \label{sec:intro}

The most agreed successful theoretical model of disc accretion was by \citet{Shakura1973}.
The most crucial result of their model for a disc around a blackhole set a condition for an accretion disc to be 
thin, i.e., the vertical scale height $(H)$ should be much less than its radial $(R)$ length scale.
Thus, the radial component of the pressure gradient is small relative to the stellar radial gravity, and the angular 
velocity is Keplerian. Also, viscosity was the main mechanism for angular momentum transfer. On the other hand angular 
momentum removal can be magnetic in origin \citep{Ghosh1978}. 

Accretion discs around magnetized stars  greatly influence the stellar magnetic field and this can result in outward 
angular momentum transfer. \cite{Ghosh1979a} presented a detailed model describing the interaction of disc and stellar 
magnetic field. They pointed out that turbulent motion, reconnection and Kelvin-Helmholtz instability allow the stellar 
magnetic field to penetrate the disc and regulate the spin of the star \citep{Ghosh1979b}. 
In fact \cite{Ghosh1979a,Ghosh1979b} found out that beyond the corotation radius, the star is spun down and vice versa. 
This is because of the impact of a slowly rotating outer part of the accretion disc. 

The presence of an intrinsic magnetic field in the accretion disc can enhance the torque acting between an accretion 
disc and an accreting star \citep{Torkelsson1998}. \cite{Tessema2010} found a complete solution of a disc structure when 
the dynamo is included. Their results show that the magnetic field that is produced by the dynamo leads to a 
significant enhancement of the magnetic torque between the neutron star and the accretion disc, compared to the model 
by \citep{Ghosh1979a,Ghosh1979b}. However, they excluded the effect of pressure gradient force (PGF). 

Inclusion of the PGF would require a slight deviation from the Keplerian motion \citep{Yi1995}.
This transition results from the internal pressure $(\sim \rho c^2_s)$ becoming a significant fraction of the orbital 
energy. The disc temperatures will be elevated above the values of an un pertubed disc \citep{Campbell1998}. 
Thus, a hot-optically thin accretion disc cannot be continously geometrically thin. In this case the vertical height 
$H\sim c_s/\Omega_k$ ($\Omega_k$ is the Keplerian angular velocity) implying that $H/R\leq1$ as opposed to $H/R\ll1$.
This is a unique feature of quasi-Keplerian rotation, in that, when the Keplerian radial distances are shifted, the 
quasi-Keplerian corotation radius as well as the position of the disc inner edge are shifted inwards \citep{Yi1997}. 

Consequently, the quasi-Keplerian model, may have observable and theoretical interesting results. 
\cite{Hoshi1977} considered quasi-Keplerian model but they never got a complete structure of the accretion 
disc. Later, \cite{Yi1997}, assumed the deviation from Keplerian fashion to be $0.2$ and found out that changes in 
magnetic torques have a visible change of spin-up or spin-down torque between the disc and neutron star. 

In this paper we seek to find a complete structure of a quasi-Keplerian, dynamo powered accretion disc around magnetized 
slowly rotating neutron star. This model follows the assumptions of \cite{Shakura1973} i.e 
Vertical hydrostatic equilibrium, steady state and $\alpha$-parameter for viscosity.
% % % % % % % % % % % % 
Then taking up the magnetized compact object model of \cite{Wang1987,Wang1995} we modify the \cite{Hoshi1977} model using 
the formulation of \cite{Tessema2010}, while taking into account the effect of radial pressure gradients. 
We subject our results to the observed data in order to explain such observational scenarios like those in 4U 
1728-247 and 4U 1626-67 \citep{Camero2010}. 

The rest of this paper is structured as follows: Section 2 presents our basic formualation, results (both theoretical 
and numerical) are discussed in Sec 3 and finally a conclusion of our findings is presented in Sec 4.
\section{Dynamical equations}
\subsection{Model description} 
The structure of the disc can be best described if we employ a cylindrical system of coordinate $(R, \phi, z)$ with 
the z-axis chosen as the axis of rotation of the 
neutron star. We consider an optically thick, geometrically thin, axisymmetric $(\partial/\partial \phi=0)$ accretion 
disc in steady state $(\partial/\partial t=0)$, taking into account the pressure gradient term and the deviation from 
Keplerian motion for a gas dominated region of the disc. 

In order to study a quasi-Keplerian accretion disc, we introduce a dimensionless variables $\xi$ showing a 
deviation from the Keplerian fashion. We assume that the azimuthal velocity is nearly Keplerian and as a result, 
values 
of $\xi$ around unity \citep{Campbell1987} such that $0<\xi \leq 2$ are considered. In the event that $\xi=1$ we regain 
the 
Keplerian form. 
The azimuthal velocity ($v_\phi$) can be modified to 
\begin{equation}
 v_\phi= 
\xi\sqrt{\frac{GM}{R}},\label{quasi}
\end{equation}
where $G$ 
is the Newton's gravitational constant, $M$ is the mass of the central object and $R$ is the radius. 

\subsection{Basic equations}
The basic equations describing the fluid dynamics in a disc are conservation of mass, momentum and energy written as:
\begin{equation}
\nabla\cdot (\rho {\bf{v}})=0,\label{b1}
\end{equation}
\begin{equation}
\rho{\bf{v}}\cdot \nabla {\bf{v}}=-\nabla P-\rho \nabla\Phi+({\bf{J}}\times {\bf{B}}) +\rho \nu 
\nabla^2{\bf{v}},\label{b2}
\end{equation}
\begin{equation}
\nabla \cdot [(\rho \varsigma +P)\cdot {\bf{v}}]={\bf{v}}\cdot{\bf{f}}_\nu -\nabla \cdot{\bf{F_{rad}}}+\frac{{\bf{J}}^2} 
{\sigma}-\nabla \cdot{\bf{q}},\label{b3}
\end{equation}
respectively. 
Here $\rho$ is the density, ${\bf{v}}=(v_R,v_\phi, v_z)$ is the fluid velocity, $\Phi=GM(R^2+z^2)^{-1/2}$ is the 
gravitational potential of the central object, $ {\bf{J}}=(J_R,J_\phi, J_z) ~\&~ {\bf{B}}=(B_R,B_\phi, B_z)$ are the 
current density $\&$ magnetic field with the radial, azimuthal and vertical components respectively, $\varsigma$ is the 
internal energy, ${\bf{f}}_\nu$ the viscous force, ${\bf{F_{rad}}}$ is radiative energy flux and $\nu$ is the 
kinematic viscosity. The $\alpha$-prescription for viscosity was assumed to be \citep{Shakura1973},
\begin{equation}
\nu=\alpha_{ss}c_s H, \label{nu}
\end{equation}
where $\alpha_{ss}$ is a constant showing the strength of viscosity and $c_s=(P/\rho)^{1/2}$ is the sound speed.
In Eq. (\ref{b3}), the term $\frac{{\bf{J}}^2}{\sigma}$ is ohmic dissipation and the term $\nabla\cdot{\bf{q}}$ is 
heat conduction. In a quasi-Keplerian motion, the energy balance equation of \cite{Frank2002} is modified to give a 
relation between temperature and radial distance along the disc as, 
\begin{equation}
 \frac{9}{8}\xi^2\nu\Sigma\frac{GM}{R^3}=\frac{4}{3}\frac{\sigma T^4_c}{\tau},\label{T1}
\end{equation}
where $\sigma$ is the Stefan Boltzmann constant, $\Sigma$ is the surface density, $T_c$ is the 
temperature at the mid-plane of the disc and $\tau$ is the optical depth of the disc defined using free-free 
opacity given by Krammers law: 
\begin{equation}
 \tau=\frac{1}{2}\Sigma \kappa,
\end{equation}
where red $\kappa$ is the Rosseland mean opacity, given by $\kappa=\kappa_0\rho 
T_c^{-7/2}$m$^2$kg$^{-2}$K$^{-7/2}$ 
with $\kappa_0=5\times10^{20}.$
\subsection{Ansatz for magnetic field}
Properties of electromagnetic fields around magnetized, rotating neutron stars have been studied both
theoretically e.g \citep{Rezzolla2001,Bakala2010,Petri2013,Petri2014} and observationally e.g \citep{Bildsten1997}. 
\cite{Rezzolla2001} derived exact general relativistic expressions for the electromagnetic field in the exterior of a 
rotating neutron star in the approximation of a slow rotation case. They considered a misaligned dipolar stellar magnetic 
field but never determined the magnetic torques exerted onto the neutron star.
The knowledge of these properties (e.g length scales, field strength, etc) benefits in understanding several 
astrophysical situations of how the neutron 
star's magnetosphere interacts with the accretion disc. 
 \cite{Lai1999} considered a non-relativistic but misaligned dipolar magnetic field and found out that the inner 
region of the accretion disk interacting with the inclined magnetic dipole field is 
subjected to magnetic torques that induce warping and precession of the disk.  
% Recently, Petri (2013) developed a formalism to compute semi-analytically this solution in general relativity.
The \cite{Lai1999} model is the opposite of \cite{Wang1987} in terms of steller field alignment and 
rotation axis. As mentioned in Section 1, we consider a non relativistic and not tilted case for a slowly rotating 
magnetized neutron star model of \cite{Wang1987,Wang1995} and then include a dynamo action of \cite{Tessema2010} as we 
extend it to a quasi-Keplerian formulation.

In X-ray binary system, it is difficult for imposed magnetic fields to be compressed to field strengths that are 
large enough to be dynamically significant in the main part of the disc \citep{Campbell1987}. Consequently, turbulent 
dynamo action in accretion discs are vital in generating 
the required magnetic fields \citep{Brandenburg1995}. In presence of a dynamo mechanism, the
stellar field penetrates the disc and a large scale toroidal field is created with two components:(1) $B_{\phi, 
\textrm{shear}}$, due to vertical shearing motions \citep{Wang1987} and (2) $B_{\phi, \textrm{dyn}}$ which is due to 
differential rotation \citep{Brandenburg1995}. 
The vertical field component, $B_{z, \textrm{dipole}}$ is assumed to take the form \citep{Wang1995}
% of the magnetic field has two parts; $B_{z, \textrm{dipole}}$ and $B_{z, \textrm{dyn}}$. 
% $B_{z, \textrm{dipole}}$ is the stellar dipolar magnetic field 
\begin{equation}
 B_{z, \textrm{dipole}}=-\frac{\mu}{R^3}, \label{Bz}
\end{equation}
where $\mu$ is the magnetic dipole moment.

The Sheared component of the dipole magnetic field $B_z$ is given by
\begin{equation}
 B_{\phi, \textrm{shear}}=-\gamma B_z \left[ 1-\left(\frac{\Omega_s}{\Omega'_k}\right)\right], \label{shear}
\end{equation}
where $\Omega'_k=\xi v_\phi/R$ is the angular velocity of the quasi-Keplerian disc, $\Omega'_k$ and $\Omega_k$ 
are related in a way that 
$\Omega'_k=\xi \Omega_k(R)<{\textrm{or}}>\Omega_k(R)$ depending on the value of $\xi$. 
We can consider the relation of $\Omega'_k$ 
and $\Omega_k$ as: 

\begin{equation} \frac{\Omega'_k}{\Omega_k} = 
\left\{ \begin{array}{lr}
\xi=1; & \textnormal{for Keplerian}, \\
0<\xi<2; & \textnormal{for quasi-Keplerian motion.} \end{array} \right.
\end{equation}
In Equation (\ref{shear}), $\Omega_s$ is the angular velocity of the star while $\gamma \gtrsim1$ \citep{Ghosh1979a} is a 
dimensionless parameter defined as the ratio of radial distance $R$ to the vertical velocity shear length 
scale $|v_\phi/(\partial v_\phi/\partial z )|$ \citep{Yi1995}. 
In this case, $\gamma$ depends on the steepness of the vertical ($z-$direction) transition between the quasi-Keplerian 
motion inside the disc and quasi-Keplerian corotation with the star outside the accretion disc i.e, 
\begin{equation}
 B_{\phi, \textrm{shear}}=\frac{\gamma \mu}{R^3} \left[1-\frac{1}{\xi}\left(\frac{R}{R'_{co}}\right)^{3/2}\right]. 
\label{Bs}
\end{equation}
In this model $R'_{co}=\xi^{2/3}R_{co}$ is the quasi-Keplerian corotation radius where $R_{co}$ is the usual corotation 
radius expressed as \citep{Tessema2010},
\begin{equation}
 R_{co}=\left(\frac{GM P^2_{{\textrm{spin}}}}{4\pi^2}\right)^\frac{1}{3}=1.5\times 10^6 
 P_{{\textrm{spin}}}^{\frac{2}{3}} M_1^{\frac{1}{3}}, 
\end{equation}
where $P_{{\textrm{spin}}}=2\pi/\Omega_s$ is the spin period of the star,  $M_1$ is the ratio $M/M_\odot$ where $M_\odot$ 
the 
solar mass. 

On the other hand, $B_{\phi, \textrm{dyn}}$ arising due to dynamo action is expressed as \citep{Tessema2011} 
\begin{equation}
 B_{\phi, \textrm{dyn}}=\epsilon(\alpha_{\textrm{ss}}\mu_0\gamma_\textrm{dyn}P(R))^{\frac{1}{2}}, \label{Bd}
\end{equation}
where $\epsilon$ is a factor which describes the direction of the magnetic field, $\mu_0$ is the 
permeability of free space and $\gamma_\textrm{dyn}=B_\phi/B_R\sim B_\phi/B_z$ \citep{Torkelsson1998} is the azimuthal 
pitch. $\gamma_\textrm{dyn}$ 
signifies the rate of 
recconection and amplification of toroidal field \citep{Campbell1999}.
% % % % % % % % % % 
% 
Here, $\alpha_{ss}~\rm{and}~ \gamma_\textrm{dyn}$ are $0.01$ \citep{Shakura1973}  and $10$ \citep{Brandenburg1995} 
respectively while $ -1 \leq \epsilon \leq +1$ . 
% % % 
The negative value shows a magnetic field which is pointing in 
the negative $\phi$ direction at the upper disc surface. 

Finally, the radial field component is given by \citep{Lai1998}
\begin{equation}
B_{R}=-\frac{B_z}{R}\left(\frac{v_R}{\Omega'_k}\right)\label{Br}
\end{equation}
which cleary depends on a quasi-Keplerian formulation $\Omega_k'$.
%
% % % % % % % % % % %  
\subsection{Disc structure}
The disc structure is fully described by the parameters: pressure, height, density, temperature and magnetic fields, 
which are obtained from simplifying the basic equations (Eq. (\ref{b1}) and (\ref{b2})).

The radial component of Eq. (\ref{b1}) is expressed as: 
\begin{equation}
\frac{1}{R}\frac{\partial}{\partial R} (\rho R v_R)+ \frac{\partial}{\partial z}(\rho v_z)=0.\label{b1a}
\end{equation}
Neglecting vertical out flows, radial integration of Eq. (\ref{b1a}), with the appropriate boundary value gives, 
\begin{equation}
 \int_{-H}^{+H}\int_{0}^{2\pi}\frac{1}{R}\frac{\partial}{\partial R} (\rho R v_R)dz d\phi=0,
\end{equation}
leading to the expression for the accretion rate
\begin{equation}
\dot{M}= -2\pi R \Sigma v_R =\textrm{constant},\label{b1mdot}
\end{equation}
where the negative sign shows inflow of matter and $\Sigma=\int_{-H}^{+H} \rho dz=2\rho H$.

Following the works of \cite{Tessema2010}, the three components of Eq. (\ref{b2}) are;  radial, 
\begin{equation}
% \begin{split}
   \rho\left[v_R\frac{\partial v_R}{\partial R}- \frac{v_\phi^2}{R}\right]  =  -\frac{\partial }{\partial R} \left[P 
+\frac{\rho GM}{(R^2+z^2)^{\frac{1}{2}}}\right]+
\left[\frac{B_z}{\mu_0}\frac{\partial B_R}{\partial z}\right]-\left[\frac{B_z}{\mu_0}\frac{\partial B_z}{\partial 
R}\right]-\left[\frac{B_\phi}{\mu_0}\frac{1}{R}\frac{\partial (RB_\phi)}{\partial R}\right],\label{b2ra}
% \end{split}
\end{equation}
azimuthal
\begin{equation}
% \begin{split}
\rho\left[v_R\frac{\partial v_\phi}{\partial R}+ \frac{v_\phi}{R}\right]=
\left[\frac{B_R}{\mu_0}\frac{1}{R}\frac{\partial (RB_\phi)}{\partial R}\right]
+
\left[\frac{B_z}{\mu_0}\frac{\partial B_\phi}{\partial z}\right]  +\frac{1}{R^2}\frac{\partial}{\partial 
R}\left[R^3\rho\nu\frac{\partial}{\partial 
R}\left(\frac{v_\phi}{R}\right)\right],\label{b2az}
% \end{split}
\end{equation}
 and vertical
\begin{equation}
% \begin{split}
  \rho\left[v_R\frac{\partial v_z}{\partial R}+ v_z\frac{\partial v_z}{\partial z}\right]  =
  - \frac{\partial }{\partial z} \left[P +\frac{\rho GM}{(R^2+z^2)^{\frac{1}{2}}}\right]
+\left[\frac{B_R}{\mu_0}\frac{\partial B_z}{\partial R}\right]-\left[\frac{B_\phi}{\mu_0}\frac{\partial 
B_\phi}{\partial z}\right]-
\left[\frac{B_R}{\mu_0}\frac{\partial B_R}{\partial z}\right].\label{b2ve}
% \end{split}
\end{equation}
From Eq. (\ref{b2ve}), vertical hydrodynamic equilibrium is expressed as: 
\begin{equation}
\frac{\partial }{\partial z}\left(P+\frac{B_R^2+B_{\phi}^2}{2\mu_0}\right)=- \rho\frac{GMz}{R^3}.\label{P_1}
\end{equation}
For a relatively high $\beta$ plasma, thermal pressure will dominate over magnetic pressure. 
On vertically integrating Eq. (\ref{P_1}) we get the pressure at the mid-plane of the disc as:
\begin{equation}
 P(R)=\frac{H\Sigma}{2}\frac{GM}{R^3}. \label{P_r}
\end{equation}
Thus, for a disc dominated by gas pressure, the equation of state for an ideal gas is 
\begin{equation}
 P(\rho, T_c)=\rho\left(
 \frac{k_B}{m_p\bar{\mu}}\right)T_c, \label{P_g}
\end{equation}
where $k_B$ is the Boltzmann constant, $m_p$, is the mass of a proton (or the mass of the hydrogen atom $m_H$, since 
$m_p\sim m_H$) and $\bar{\mu}$ is the mean molecular weight for the ionized gas. The value of $\bar{\mu}$ ranges between 
$0.5$ for fully ionized hydrogen and $1$ for neutral hydrogen (i.e $0.5\leq \bar{\mu}\leq 1$) depending on the degree of 
ionization of the gas \citep{Frank2002}. In this model we take $\bar{\mu}=0.62m_H$, which corresponds to a mixture 
of ionized gas comprised of $70\%$ hydrogen and $30\%$ helium by mass. 
Using equations, Eq. (\ref{P_r}) and Eq. (\ref{P_g}), the disc height is given as: 
\begin{equation}
H=\left(\frac{m_p\bar{\mu}}{k_B}\right)^{-\frac{1}{2}}\left(\frac{GM}{R^3}
\right)^ {-\frac{1}{2}} T_c^{\frac{1}{2}}. \label{H}
\end{equation}
From the viscous stress tensor we have, 
\begin{equation}
 \frac{3}{4}\xi (\nu\Sigma) \left(\frac{R^3}{GM}
\right)^ {-\frac{1}{2}}H^{-1}=\alpha_{\textrm{ss}} P(R).\label{P_rr}
\end{equation}
Combining Equations (\ref{P_r}), (\ref{H}) and (\ref{P_rr}) the density is given by,
\begin{equation}
\rho=\frac{3}{4}\xi(\nu\Sigma) \alpha_{\textrm{ss}}^{-1} 
\left(\frac{GM}{R^3}\right)\left(\frac{m_p\bar{\mu}}{k_B}\right)^ 
{\frac{3}{2}} T_c^{-\frac{3}{2}}.\label{De}
\end{equation}
Optical depth of the accretion disc is obtained from 
\begin{equation}
 \tau=\frac{1}{2}\Sigma \kappa=\rho^2 H \kappa_0 T_c^{-\frac{3}{2}}.\label{Tu}
\end{equation}
Substituting Eq. (\ref{H}) and Eq. (\ref{De}) into  Eq. (\ref{Tu}) the optical density is expressed as: 
\begin{equation}
 \tau=\frac{9}{16}\xi^2(\nu\Sigma)^2 \alpha_{\textrm{ss}}^{-2} \kappa_0
\left(\frac{GM}{R^3}\right)^{\frac{3}{2}}\left(\frac{m_p\bar{\mu}}{k_B}\right)^
{\frac{5}{2}} T_c^{-6}.
\end{equation}
Now the midplane temperature $T_c$ is obtained from Eq. (\ref{T1}) as: 
\begin{equation}
% \begin{split}
 T_c=\left(\frac{243\kappa_0}{512\sigma}\right)^{\frac{1}{10}} 
\left(\frac{GM}{R^3}\right)^{\frac{1}{4}}\left(\frac{m_p\bar{\mu}}{k_B}\right)^ 
{\frac{1}{4}}\alpha_{\textrm{ss}}^{-\frac{1}{5}} \xi^{\frac{2}{5}}(\nu\Sigma)^{\frac{3}{10}}.
% \end{split} 
\end{equation}
From Eq. (\ref{P_r}) we can obtain a pressure expression that is related to $R$, as: 
\begin{equation}
% \begin{split}
  P(R)=\frac{3}{4} \left(\frac{243\kappa_0}{512\sigma}\right)^{-\frac{1}{20}} 
\left(\frac{GM}{R^3}\right)^{\frac{7}{8}}\left(\frac{m_p\bar{\mu}}{k_B}\right)^ 
{\frac{3}{8}} \alpha_{\textrm{ss}}^{-\frac{9}{10}} \xi^{\frac{4}{5}}(\nu\Sigma)^{\frac{17}{20}}.
% \end{split} 
\label{P}
\end{equation}
Surface density, $\Sigma$ and radial velocity, $v_R$, take the form:
\begin{equation}
% \begin{split}
  \Sigma=\frac{3}{2}\left(\frac{243\kappa_0}{512\sigma}\right)^{-\frac{1}{10}} 
\left(\frac{GM}{R^3}\right)^{\frac{1}{4}}\left(\frac{m_p\bar{\mu}}{k_B}\right)^ 
{\frac{3}{4}}\alpha_{\textrm{ss}}^{-\frac{4}{5}}\xi^{\frac{2}{5}}(\nu\Sigma)^{\frac{7}{10}} ,
% \end{split}
\label{S}
\end{equation}
and 
\begin{equation}
% \begin{split}
  v_R=-\frac{\dot{M}}{3\pi}\left(\frac{243\kappa_0}{512\sigma}\right)^{\frac{1}{10}} 
\left(\frac{m_p\bar{\mu}}{k_B}\right)^{-\frac{3}{4}}(GM)^{-\frac{1}{4}} 
 \alpha_{\textrm{ss}}^{\frac{4}{5}}\xi^{-\frac{2}{5}}(\nu\Sigma)^{-\frac{7}{10}} R^{-\frac{1}{4}}.
%  \end{split}
 \label{S}
\end{equation}
This layout gives a basis to find a complete structure of a quasi-Keplerian accretion disc. 
% % % % % % % % % % % % % % % % % % % % % 

\section{Results and Discussion}
\subsection{Global Solutions}
All structural equations appear as a function of $\nu\Sigma$.
This can be made explicit from the azimuthal component of momentum equation (Eq. (\ref{b2az})) by integrating 
vertically which yields:
\begin{equation}
% \begin{split}
\Sigma\left[v_R\frac{\partial \ell}{\partial R}\right]=R
\left[\frac{B_z B_\phi}{\mu_0}\right]_{z=-H}^{z=+H}  
+\frac{1}{R}\frac{\partial}{\partial R} \left[R^3(\nu \Sigma)
\frac{\partial}{\partial R}\left(\frac{\ell}{R^2}\right)\right],
% \end{split}
\label{b2azint}
\end{equation}
where $\ell=R v_\phi$ is the specific angular momentum. Here we have eliminated the term in 
$\left[\frac{B_R}{\mu_0}\frac{1}{R}\frac{\partial(RB_\phi)}{\partial R}\right]$ due to spatial difference.
Taking $B_{\phi,\textrm{dyn}} B_{z,\textrm{dipole}}$ and $B_{\phi,\textrm{shear}}B_{z,\textrm{dipole}}$ as the dominant 
terms \citep{Tessema2010} of the expansion for $B_z B_\phi$ term in Eq. (\ref{b2azint}), we have 
\begin{equation}
% \begin{split}
 \Sigma\left[v_R\frac{\partial \ell}{\partial R}\right]=2R [B_z (B_{\phi, \rm{dyn}}+B_{\phi, \rm{shear}})]  
+\frac{1}{R}\frac{\partial}{\partial R} \left[R^3(\nu \Sigma) \frac{\partial}{\partial R} 
\left(\frac{\ell}{R^2}\right)\right],
% \end{split}
\label{azimuthal}
\end{equation}

Using equations (Eq. \ref{quasi}, \ref{Bz}, \ref{Bs}, \ref{Bd} and \ref{b1mdot}) in Eq. (\ref{azimuthal}) we have 
\begin{equation}
%  \begin{split}
 y'=\frac{\dot{M}}{6\pi R}\xi^{-\frac{2}{5}}-\frac{y}{2 R} -C_1  \epsilon \xi^{-\frac{3}{5}}  
y^{\frac{17}{40}}R^{-\frac{45}{16}} 
 - C_2  \xi^{-1}R^{-\frac{9}{2}} \left[1-\frac{1}{\xi}\left( \frac{R}{R_{co}}\right)^{\frac{3}{2}}\right],\label{deiny}
%   \end{split}
\end{equation}
where $C_1= 
\left[\left(\frac{4}{3}\frac{
\gamma_{\textrm{dyn}}}{\mu_0}\right)^{\frac{1}{2}}\left(\frac{243\kappa_0}{512\sigma}\right)^{-\frac{1
}{40}} \left(\frac{m_p\bar{\mu}}{k_B}\right)^{\frac{3}{16}}\mu \alpha_{\textrm{ss}}^{\frac{1}{20}} 
(GM)^{-\frac{1}{16}}\right]$ and $C_2=\frac{4}{3}\frac{\mu^2}{\mu_0}\gamma(GM)^{-\frac{1}{2}}$.
This is a differential equation in $y$ for the quasi-Keplerian case which is analogous to Eq. (41) of \cite{Tessema2010} 
only when the value of $\xi=1$. 

We need to transform Eq. (\ref{deiny}) by introducing dimensionless quantities; 
$\Lambda$ and $r$, so that
\begin{equation}
y=\Lambda \dot{M}\label{dimensionless1}
\end{equation}
\begin{equation}
R= r R_A.\label{dimensionless2}
\end{equation}
Here $r$ is a dimensionless radial coordinate and $R_A$ is the Alfv\'{e}n radius which is a characteristic radius at 
which magnetic stresses dominate the flow in the accretion disc. It is obtained by equating the magnetic pressure to the 
ram pressure \citep{Frank2002}. 
\begin{equation}
R_A=\left(\frac{2\pi^2 \mu^4}{GM\dot{M}^2\mu_0^2}\right)^{\frac{1}{7}}= 
5.1\times10^6\dot{M_{13}}^{-\frac{2}{7}}M_1^{-\frac{1}{7}}\mu_{20}^{\frac{ 4 } { 7 } }\rm{m},
\end{equation}
where $\dot{M}$ is the rate of accretion, $\mu_{20}$ is the steller magnetic dipole moment in units of $10^{20}$Tm$^3$ 
and $\dot{M_{13}}$ is the accretion rate in units of $10^{13}$kgs$^{-1}$.

Finally, we get a differential equation in $\Lambda$ as:
\begin{equation}
%  \begin{split}
 \Lambda'=\frac{1}{6\pi r}\xi^{-\frac{2}{5}}-\frac{\Lambda}{2 r} - C_3  \epsilon\xi^{-\frac{3}{5}}\Lambda^{\frac{17}{40}} 
r^{-\frac{45}{16}}-C_{4}\xi^{-1} r^{-\frac{9}{2}} \left[ 1 -\frac{\omega_s}{\xi} r^{\frac{3}{2}}\right],
%     \end{split}
\label{deinLa}
\end{equation}
where $C_3=C_1 \dot{M}^{\frac{17}{40}} R_A^{-\frac{45}{16}}$, $C_{4}=C_2 R_A^{-\frac{9}{16}}$ and $\omega_s$ is a 
fastness parameter defined as \citep{Elsner1977}, $\omega_s=\left(R_A/R_{co}\right)^{\frac{3}{2}}=6.3M_1^{-\frac{5}{7}} 
\dot{M}^{-\frac{3}{7}}_{13}\mu_{20}^{\frac{6}{7}}P^{-1}$. 
Equation (Eq. (\ref{deinLa})) is the new analytical solution, for a quasi-Keplerian model. 
In the limit $\omega_s<1$, steady accretion takes place while for $\omega_s>1$ accretion is unsteady and the 
accreting matter will be propelled outward by centrifugal forces. Also, we note that as $r\rightarrow\infty$ say 
$100R_A$ then $\Lambda\rightarrow 1/3\pi$, which becomes the boundary condition for this model.

In the absence of 
magnetic field ($\gamma=0$), the internal dynamo ($\gamma_\textrm{dyn}=0$) and the quasi-Keplerian assumption ($\xi=1$), 
the expression for 
$\Lambda$ reduces to
\begin{equation}
 \frac{d\Lambda}{dr}=-\frac{\Lambda}{2 r}+\frac{1}{6\pi r},
\end{equation}
which is similar to the Shakura-Sunyaev (SS) model equation, originally derived by 
\cite{Shakura1973}, in the classical model around black holes in binary system.

 Close examination 
of equation (\ref{deinLa}) shows that there are two possible boundary conditions to locate the inner edge of the disc. We 
can define them as Case D and Case V \citep{Tessema2010}. Firstly, in case D, the inner edge is located at a radial 
distance where the density and temperature drop to zero as the inflow velocity becomes infinite, 
meaning that $\Lambda=0$ \citep{Shakura1973}. Secondly, in case V, the inner edge is located at a radial 
distance where the disc plasma is driven along field lines by transfer of excess angular momentum \citep{Wang1995}. In 
this case $\Lambda \neq 0$.

In our model, we consider a neutron star which is accreting at a rate of $10^{13}$kgs$^{-1}$ with a mass 
$M=1.4M_\odot$ and a magnetic moment of $10^{20}$Tm$^{3}$. We fix the parameters $\alpha_{ss},~ \gamma~ \rm{and} 
~\gamma_{dyn}$ to $0.01, ~1,~ \rm{and} ~10$ respectively. 
Through out our working we set the quasi-Keplerian parameter $\xi$ to 0.8 \& 1.2, since azimuthal velocity is close to 
Keplerian. This will enable us observe the 
behaviour of the disc as it transits to and from Keplerian fashion with the azimuthal velocity varying by 20\% below and 
above the Keplerian azimuthal velocity. The spin periods of interest in this model are $7~\rm{and}~100s$, and for each 
spin period we obtain a solution while changing the dynamo parameter, $\epsilon=1.0,~ 0.1,~ 0, ~-0.1~ \rm{and}~ -1$. 
These spin periods are purposely chosen because they cover a wide range of slowly rotating stars that 
exhibit torque reversals, for example 4U 1626-67 (Period=7.6s) and 4U 1728-247 (Period=120s) 
\citep{Bildsten1997,Camero2010}.
% % % % % % 

On the other hand, rapid rotators such as SAX J1808.4-3658, have also been observed to have spin variations 
 \citep{Burderi2006} that can be best explained in a dynamo model \citep{Tessema2011}. They behave uniquely 
when the accreting plasma is threaded by the stellar magnetic field \citep{Naso2010,Naso2011}.  
% % % % % %  
To find a complete structure of such discs, the disc is divided into regions depending on the equation of state.
For example, the SS model solution considers the regions of the disc to have pressure either dominated
by radiation pressure or ideal gas pressure and the main opacity source is electron scattering
 or Kramer's opacity \citep{Shakura1973,Shakura1976}. However, to analyse the dynamics of a quasi-Keplerian 
accretion disc, this present article focuses on a gas pressure dominated disc around slowly rotating magnetized
neutron stars.
% % % % % % % % % % % % 
%L
\begin{figure*}
\gridline{\fig{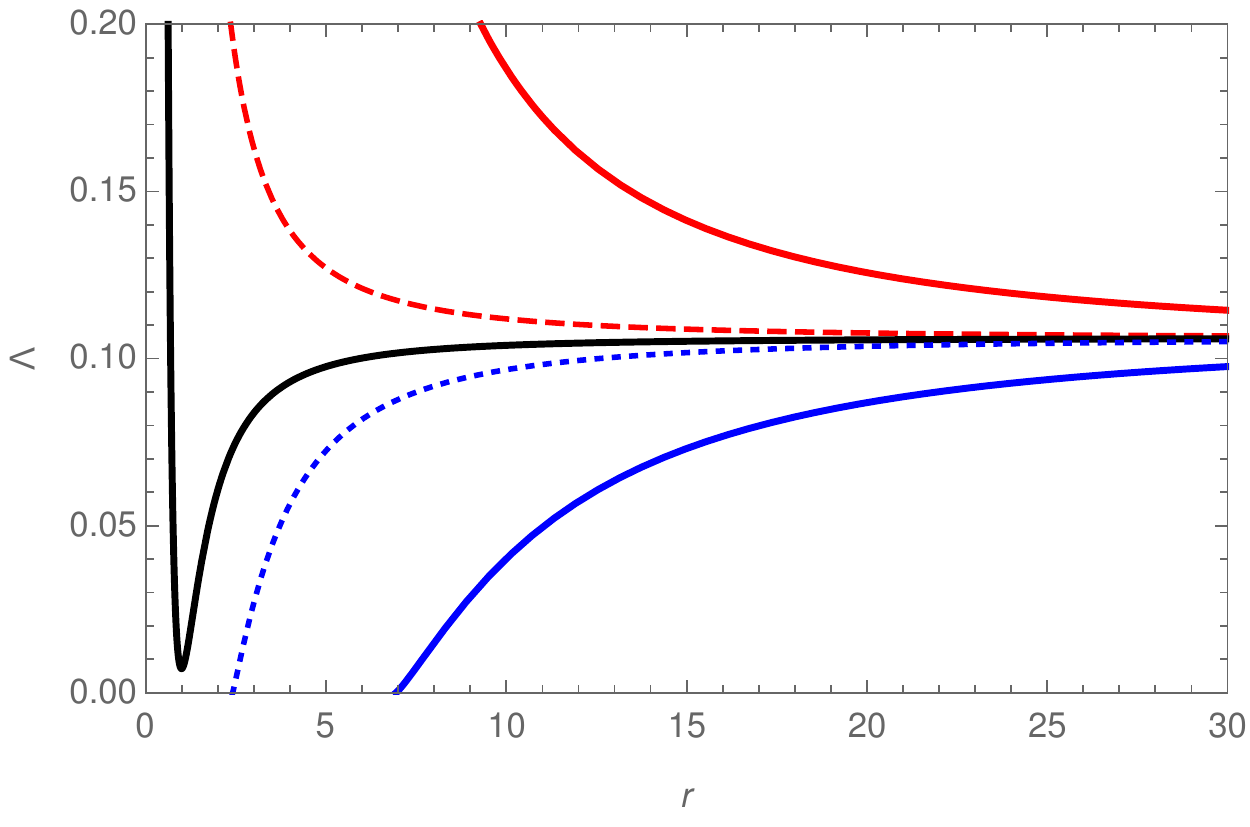}{0.3\textwidth}{(a)$\xi=0.8$}
          \fig{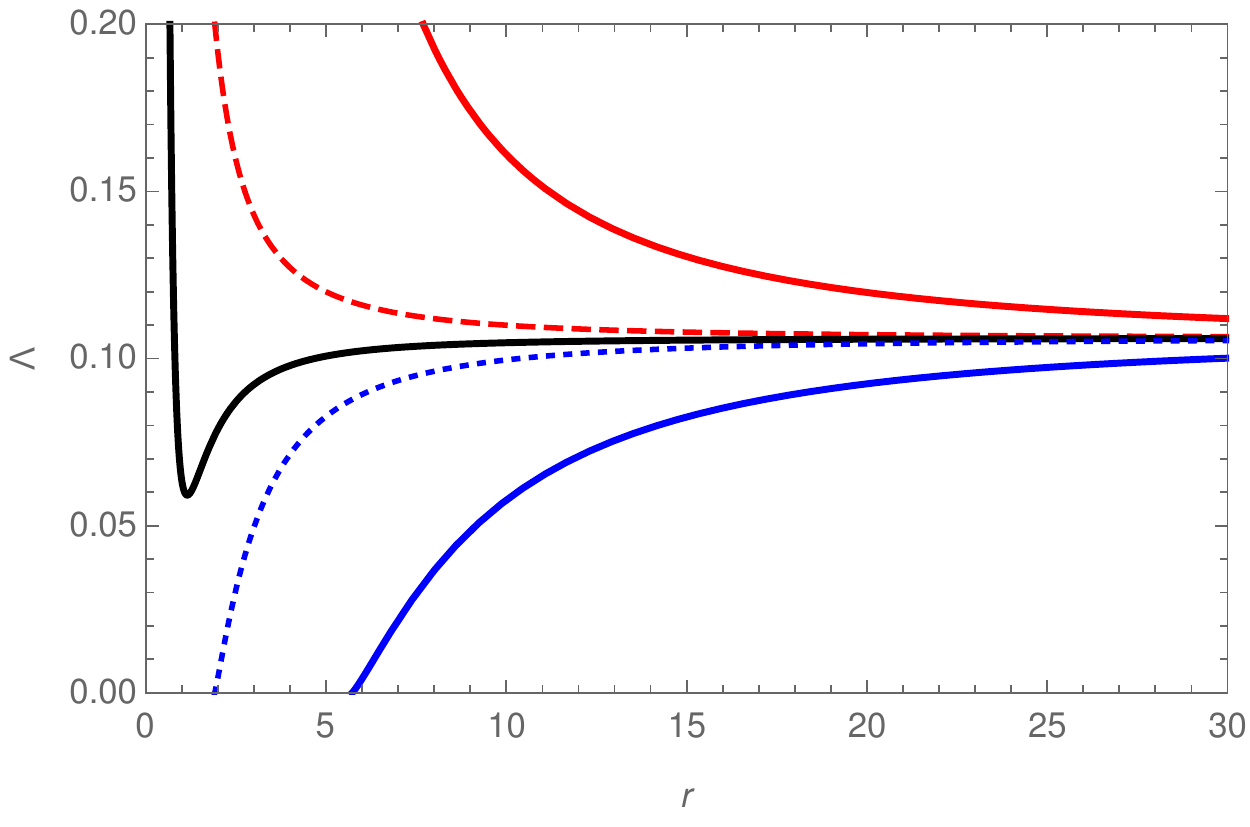}{0.3\textwidth}{(b)$\xi=1.0$}
          \fig{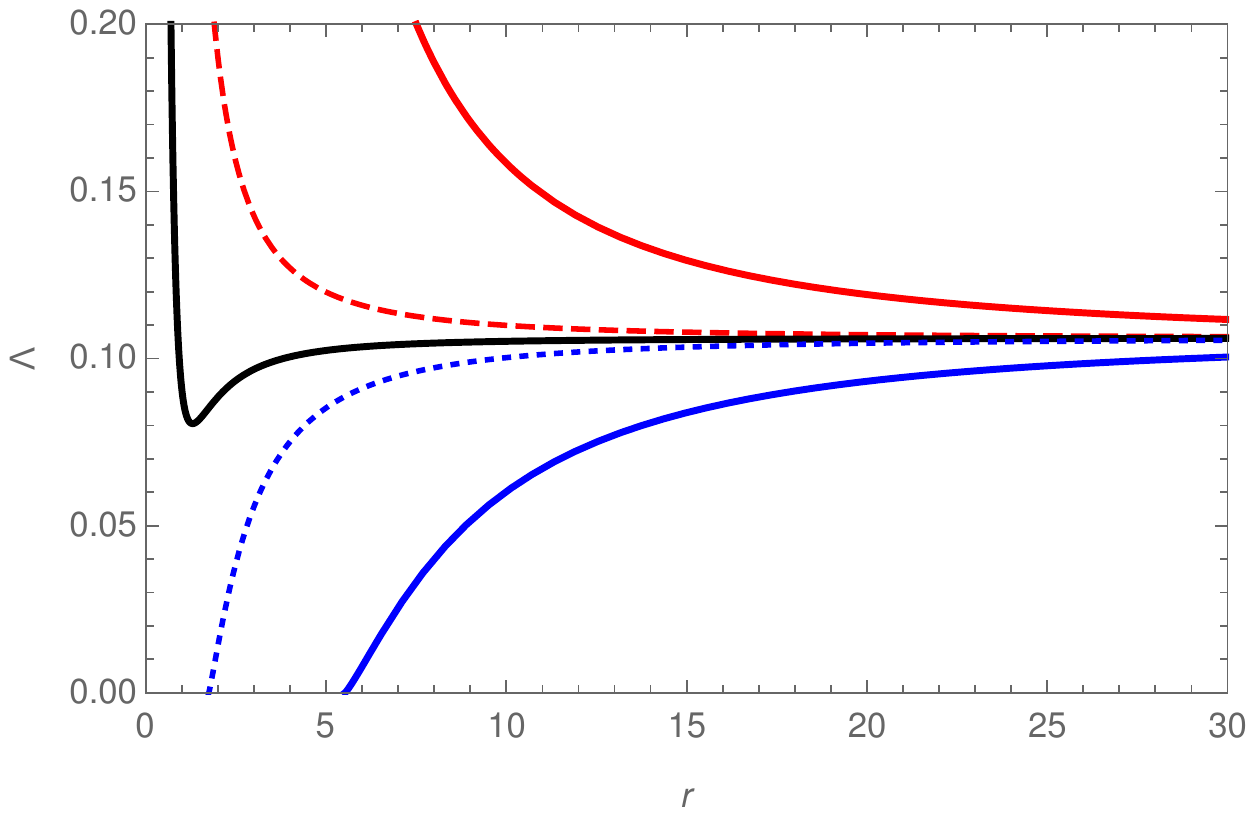}{0.3\textwidth}{(c)$\xi=1.2$}
          }
\gridline{\fig{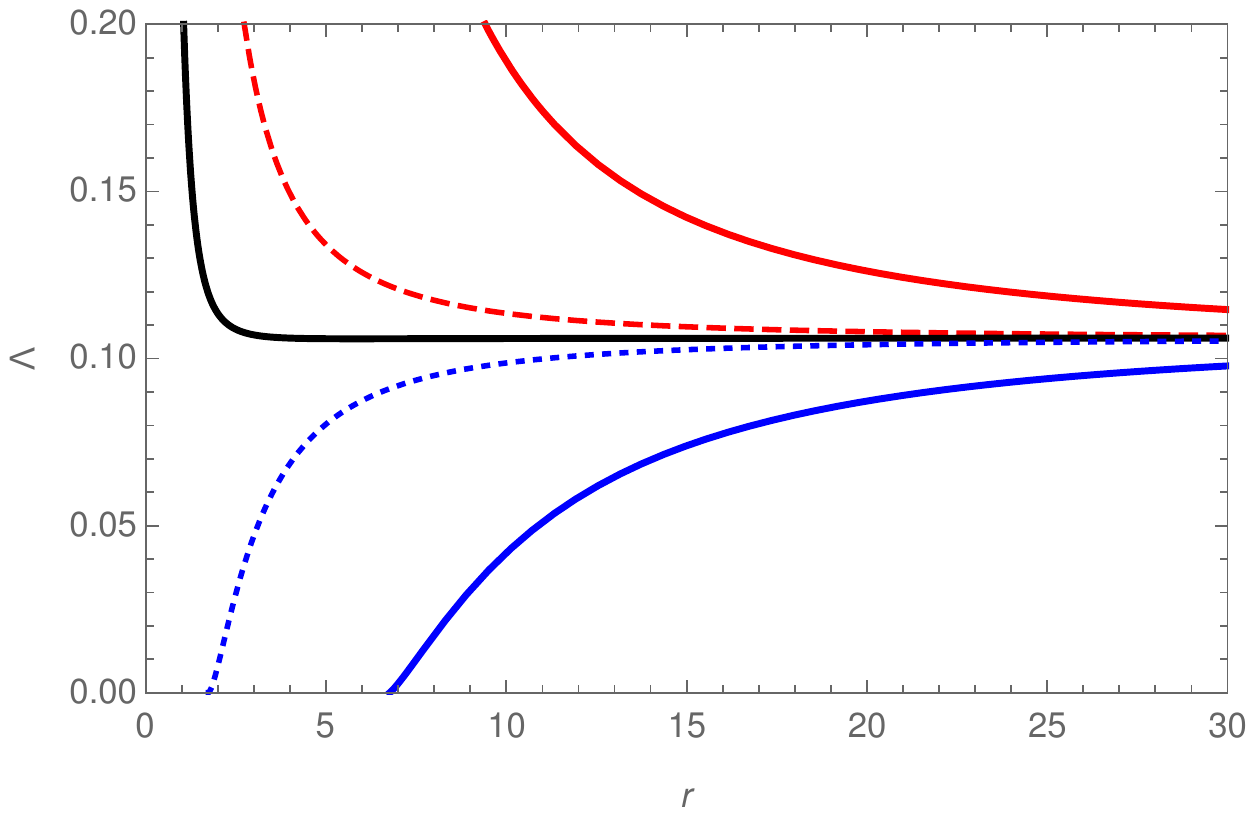}{0.3\textwidth}{(d)$\xi=0.8$}
          \fig{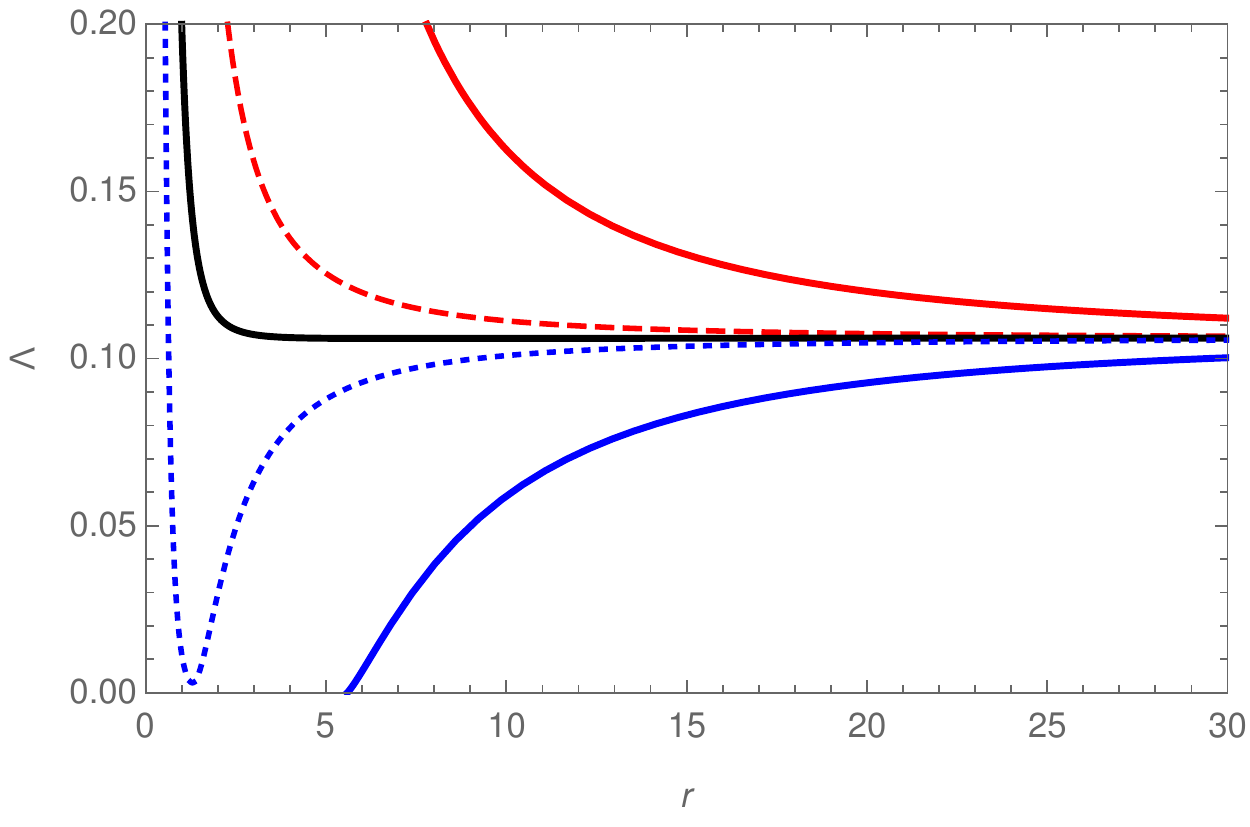}{0.3\textwidth}{(e)$\xi=1.0$}
          \fig{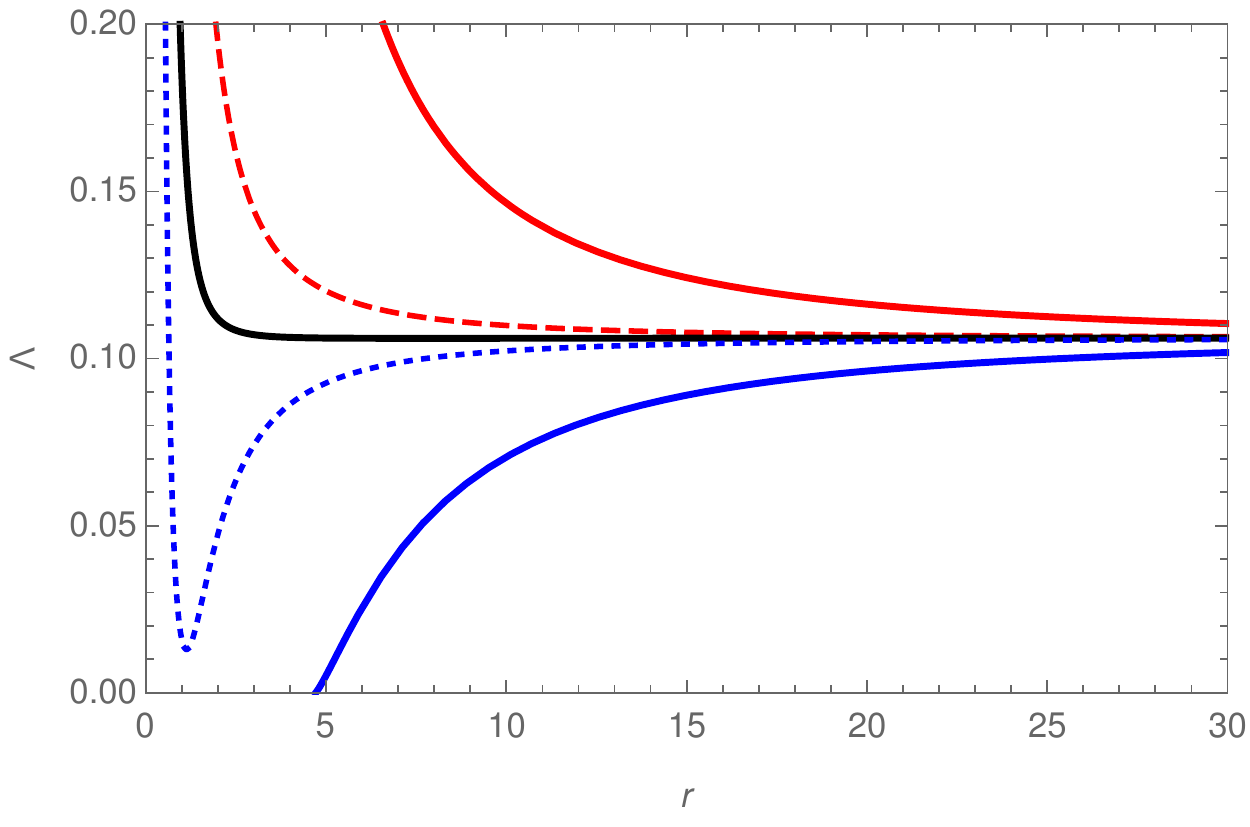}{0.3\textwidth}{(f)$\xi=1.2$}}
\caption{Variation of $\Lambda(r)$ with radial distance for a neutron star with a spin period of: 7s ($top~panel$), 100s 
 ($bottom~panel$). The magnetic field generated by the dynamo are shown with: $\epsilon=-1.0$ 
 blue thick, 
 $\epsilon=-0.1$ blue dotted, $\epsilon=0$ black, $\epsilon=0.1$ red dashed and $\epsilon=1.0$ red thick \label{Lambda}}
\end{figure*}
% % % % % % % 
 %cv
 \begin{figure*}
\gridline{\fig{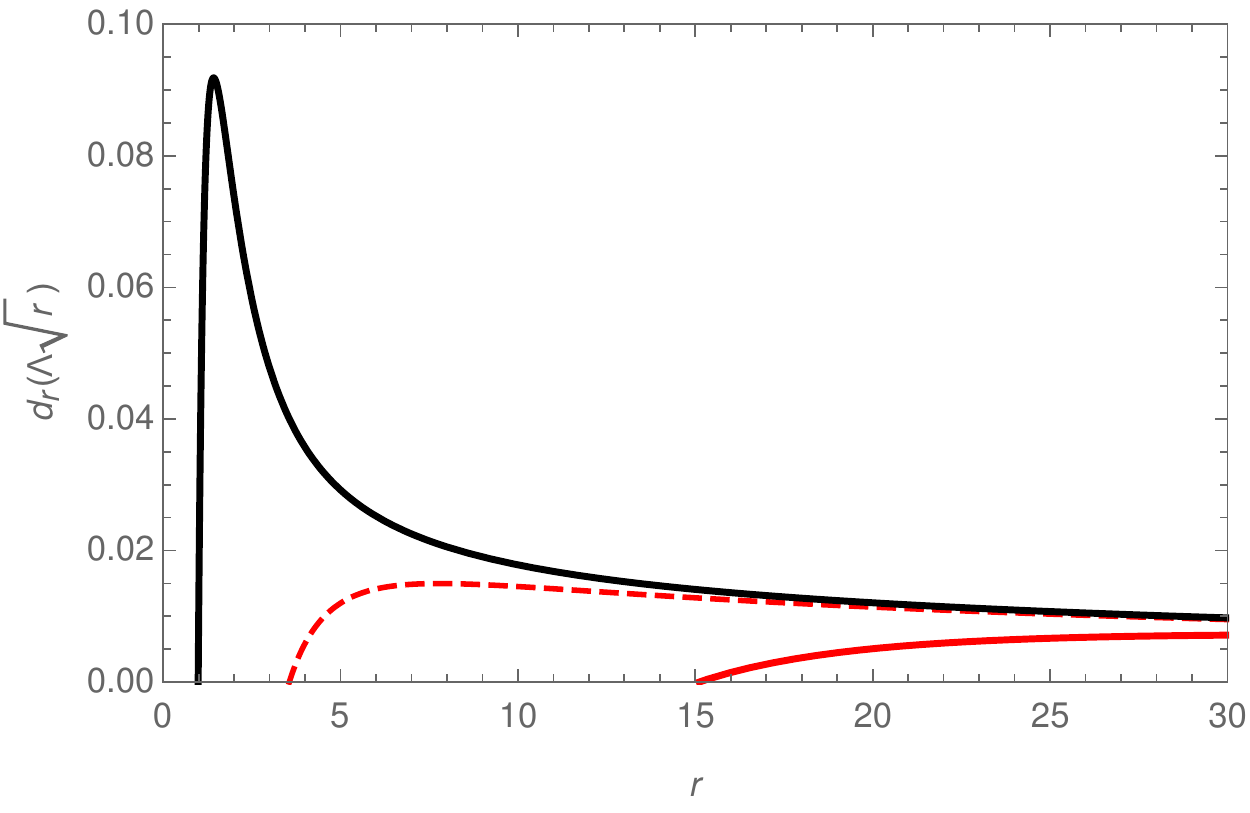}{0.3\textwidth}{(a)$\xi=0.8$}
          \fig{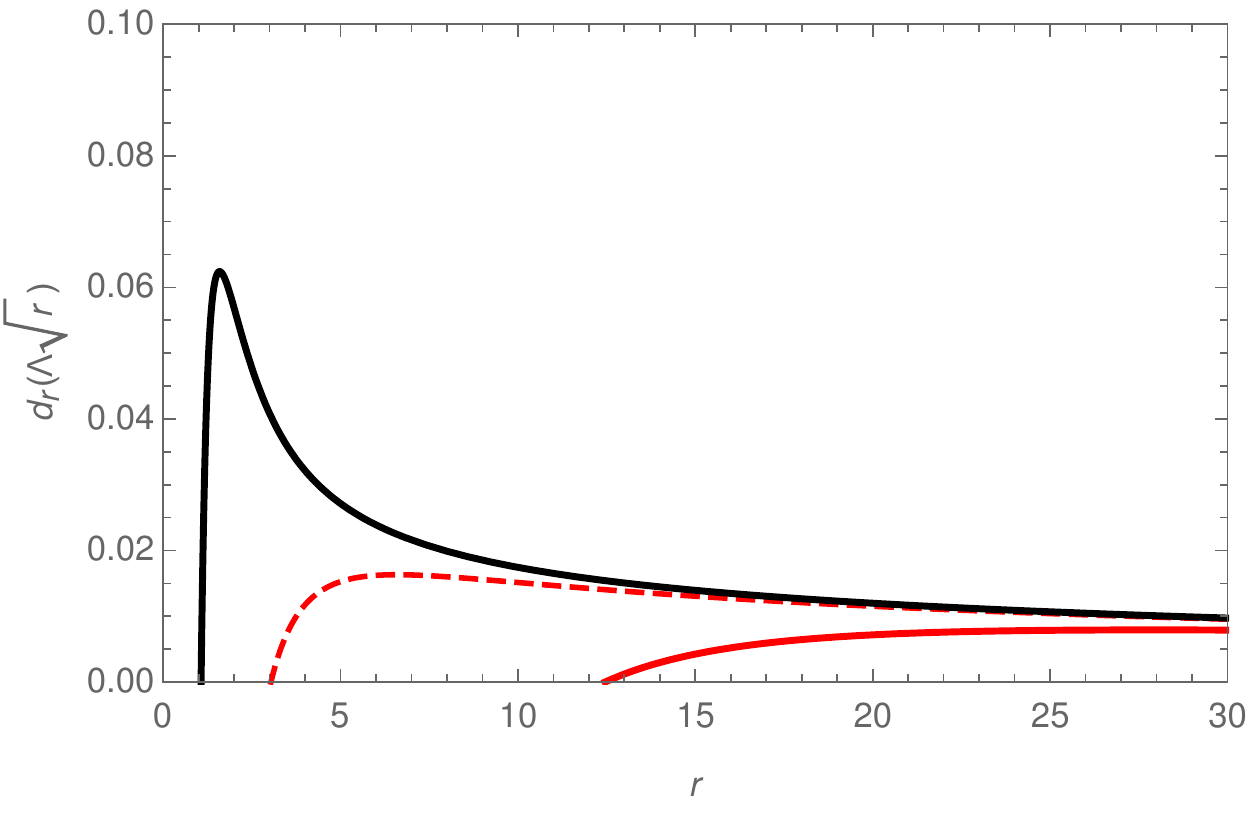}{0.3\textwidth}{(b)$\xi=1.0$}
          \fig{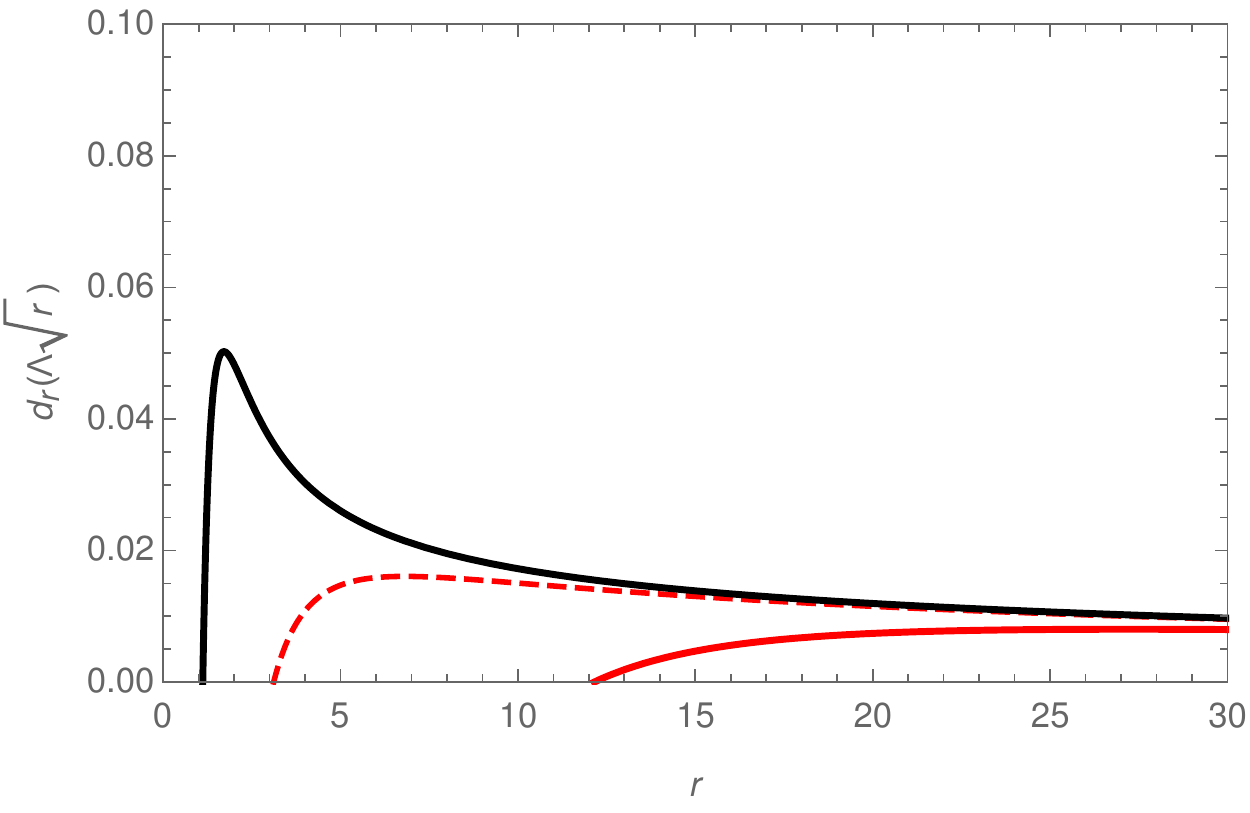}{0.3\textwidth}{(c)$\xi=1.2$}
          }
\gridline{\fig{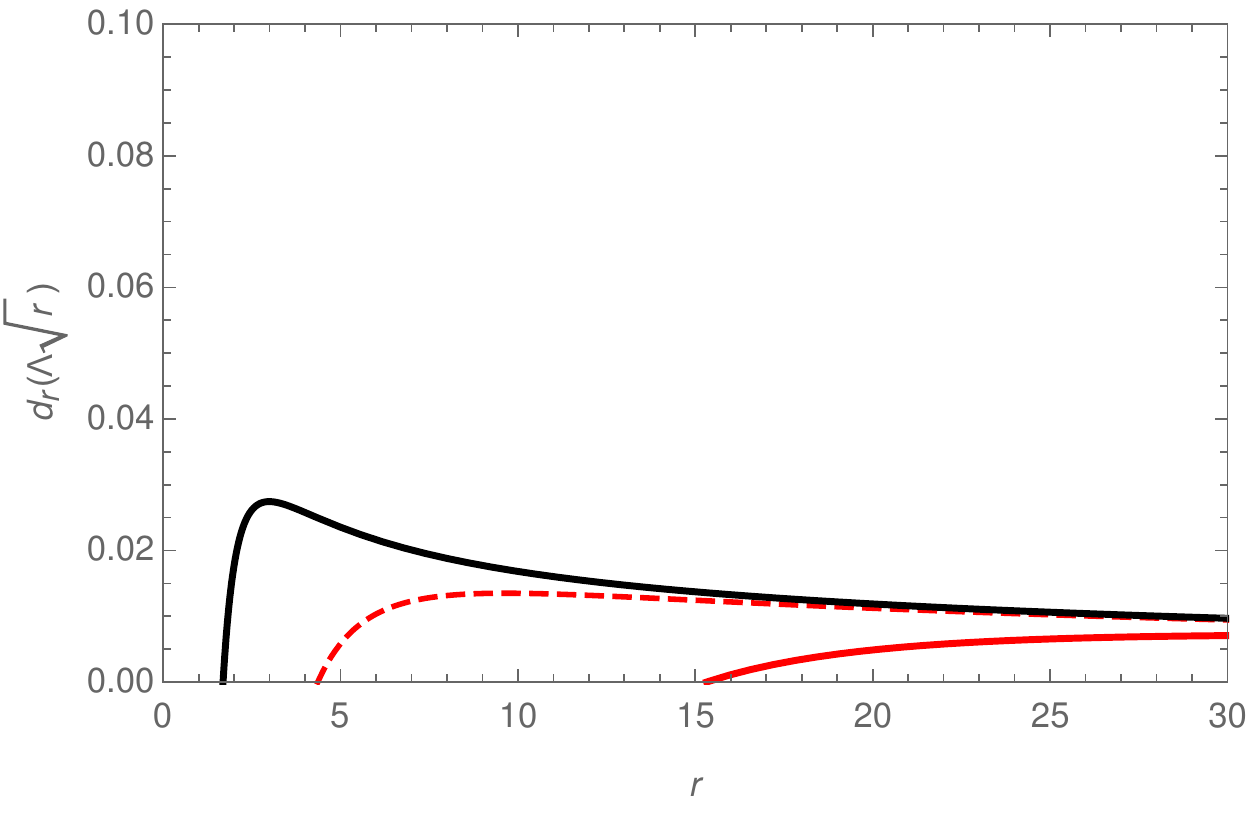}{0.3\textwidth}{(d)$\xi=0.8$}
          \fig{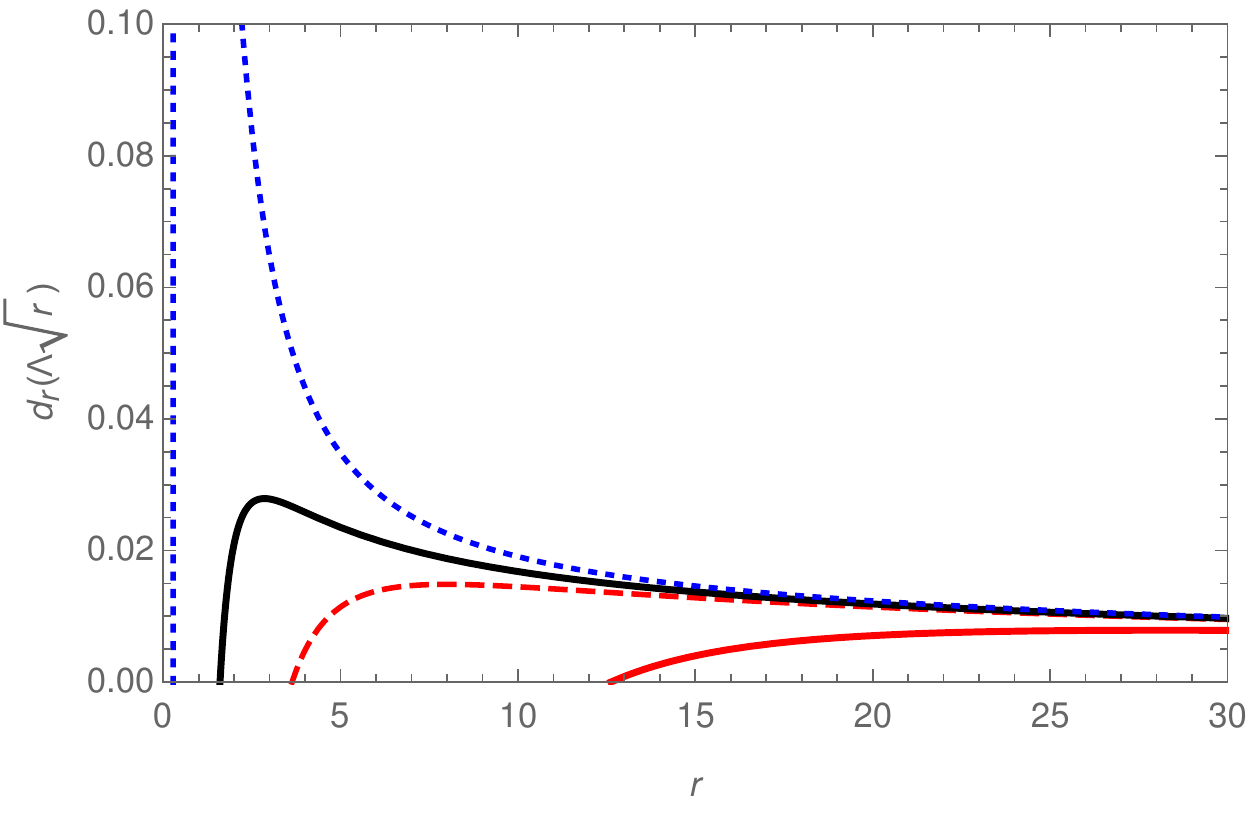}{0.3\textwidth}{(e)$\xi=1.0$}
          \fig{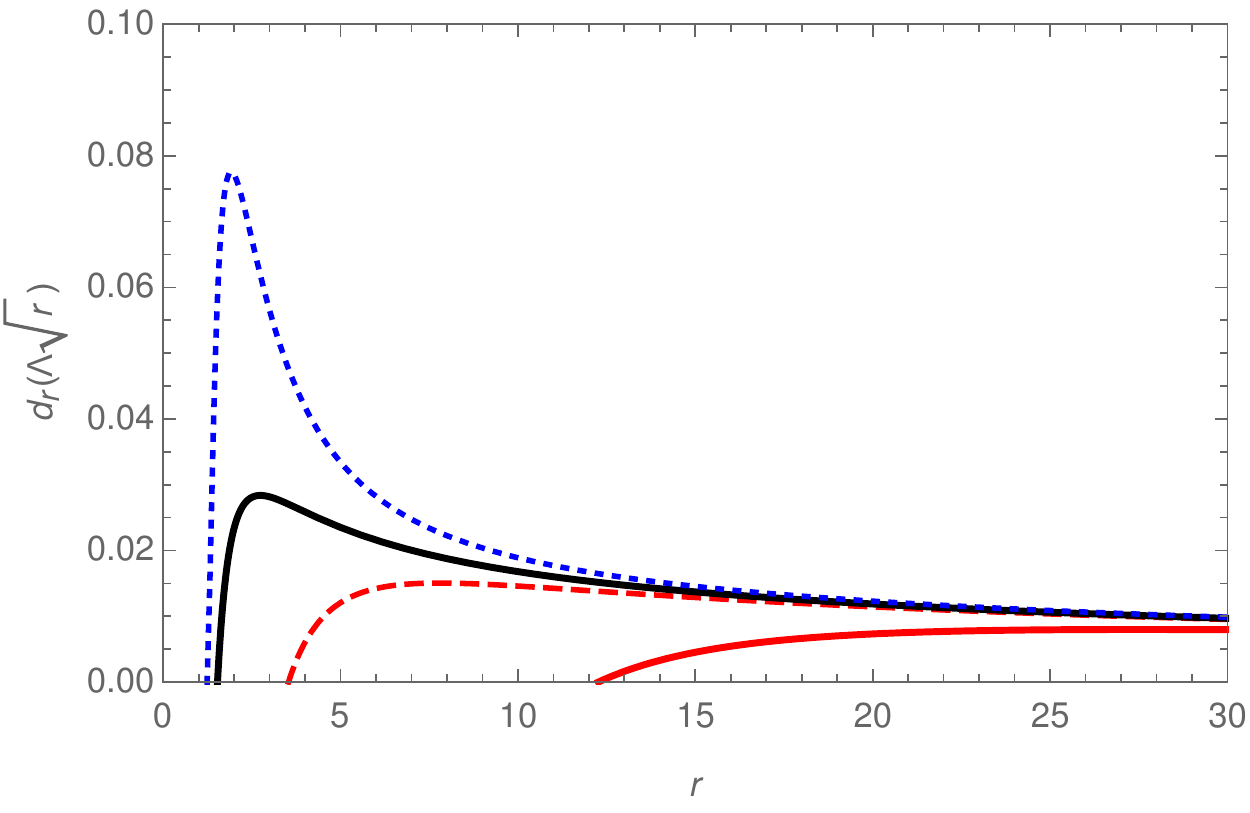}{0.3\textwidth}{(f)$\xi=1.2$}}
\caption{Variation of $\frac{d}{dr}(\sqrt{r}\Lambda(r))$ with radial distance for a neutron star with a spin period of: 
7s ($top~panel$), 100s ($bottom~panel$). The magnetic field generated by the dynamo are shown with: $\epsilon=-1.0$ 
blue thick, $\epsilon=-0.1$ blue dotted, $\epsilon=0$ black, $\epsilon=0.1$ red dashed and $\epsilon=1.0$ red 
thick}\label{cv}
\end{figure*}

The global solutions show that at large radial distance, both Keplerian and quasi-Keplerian motion exhibit nearly the 
same $\Lambda$ variations with radius (Figure \ref{Lambda}). 
As the quasi-Keplerian disc interacts with the stars magnetosphere, the corotation radius is shifted inwards for 
$\Omega'_k>\Omega_k$ and outwards for $\Omega'_k<\Omega_k$. Subsequently, the position of the inner edge 
with respect to the new corotation radius is relocated. 
When $\Omega'_k>\Omega_k>\Omega_s$ ($\Omega'_k<\Omega_k<\Omega_s$) the disc plasma is moving faster (slower) than the 
star and consequently, magnetic stresses act to spin up (down) the star \citep{Wang1995}. 
Further, we note that by varying the quasi-Keplerian parameters, the disc structure is modified and the star experiences 
enhanced torques. 
In previous studies quasi-Keplerianity has been restricted to only when $\Omega'_k<\Omega_k$ \citep{Yi1995}. This study 
explored both situations.

Figure (\ref{Lambda}) shows the variation of $\Lambda$ as a function of $r$ for both $P=7s$ and $P=100s$.
All $\epsilon=1.0, 0.1, 0$ solutions are case V inner boundaries in addition to $\epsilon=-0.1$ for $P=100s$ when 
$\xi=1.0$ and $\xi=1.2,$ see Figure (\ref{cv}).
For $P=7s$, the local minimum that occurs when $\epsilon=0$  keeps disapearing as the disc gets 
into quasi-Keplerian state (Figure \ref{Lambda}) $top~ pannel$. 
In this transition, the dynamo plays a big role on the nature of the global solution.
Dynamo action was found to results in enhanced magnetic torques between the star and disc \citep{Tessema2010}.  
Therefore, a combination of dynamo action and quasi-Keplerian situation has an effect on torque reversal.
We beleive that this is a possible 
physical situation that arise in an accretion disc at such inner radius close to the neutron star.

\subsection{The structure of a quasi-Keplerian disc}
The structural equations are obtained by expressing all the unknowns  $P,T_c, H/R, \Sigma, \rho, \tau, \nu, v_R, 
B_{\phi,\textrm{dyn}},$ and $B_{R}$ in terms of $\alpha_{\textrm{ss}}$, $\dot{M_{13}}$, $M_1, 
\mu_{20}, \bar{\mu}, \Lambda(r)$ and $r$ obtained as: 
\begin{equation}
\Sigma=3.8\times 10^3 
\alpha_{\textrm{ss}}^{-\frac{4}{5}}\xi^{\frac{2}{5}}\bar{\mu}^{\frac{3}{4}}\mu_{20}^{-\frac{3}{7}}\dot{M_{13}}^{\frac{32}{
 35 } }
 M_1^{\frac{5} {14}} \Lambda(r)^ { \frac{7}{10}} r^{-\frac{ 3}{4}} \rm{kgm}^{-2}
\end{equation}
\begin{equation}
\rho=3.0\times 
10^{-2}\alpha_{\textrm{ss}}^{-\frac{7}{10}}\xi^{1}\bar{\mu}^{\frac{9}{8}}\mu_{20}^{-\frac{15}{14}}\dot{M_{13}}^{
\frac{
38}{35}} M_1^{\frac{25} {28}} \Lambda(r)^ {\frac{11}{20}} r^{-\frac{15}{8}}\rm{kgm}^{-3}
\end{equation}
\begin{equation}
 P=1.2\times 
10^8\alpha_{\textrm{ss}}^{-\frac{9}{10}}\xi^{\frac{4}{5}}\bar{\mu}^{\frac{3}{8}}\mu_{20}^{-\frac{3}{2}}\dot{M_{13}}^{\frac
{ 8}{5}} 
M_1^{\frac{5} {4}}\Lambda(r)^ {\frac{17}{20}}  r^{-\frac{ 21}{8}} \rm{Nm}^{-2}
\end{equation}
\begin{equation}
\nu=4.0\times10^9 
\alpha_{\textrm{ss}}^{\frac{4}{5}}\xi^{\frac{1}{5}}\bar{\mu}^{-\frac{3}{4}}\mu_{20}^{\frac{3}{7}}\dot{M_{13}}^
{\frac{3}{35}} M_1^{-\frac{5} {14}}\Lambda(r)^ {\frac{3}{10}} r^{\frac{3}{4}} \rm{m^2s}^{-1}
\end{equation}
\begin{equation}
 T_c=4.8\times 10^5 \alpha_{\textrm{ss}}^{-\frac{1}{5}}\xi^{\frac{2}{5}}\bar{\mu}^{\frac{1}{4}}\mu_{20}^{-\frac{3}{7}} 
\dot{M_{13}}^{\frac{18}{35}} M_1^{\frac{5} {14}} \Lambda(r)^ {\frac{3}{10}} r^{-\frac{3}{4}} \rm{K}
\end{equation}
\begin{equation}
\frac{H}{R}=1.2\times 10^{-2} 
\alpha_{\textrm{ss}}^{-\frac{1}{10}}\xi^{\frac{1}{5}}\bar{\mu}^{-\frac{3}{8}}\mu_{20}^{\frac{1}{14}} 
\dot{M_{13}}^{-\frac {4}{35}} M_1^{-\frac{11} {28}}\Lambda(r)^ {\frac{3}{20}}r^{\frac{1}{8}} 
\end{equation}
\begin{equation}
v_R=82.8\alpha_{\textrm{ss}}^{\frac{4}{5}}\xi^{-\frac{2}{5}}\bar{\mu}^{-\frac{3}{4}}\mu_{20}^{-\frac{1}{7}}\dot{M_{13}}^{
\frac{13}{35}} 
M_1^{-\frac{3} {14}}\Lambda(r)^ { -\frac{7}{10}} r^{-\frac{ 1}{4}}  \rm{ms}^{-1}
\end{equation}
\begin{equation}
\tau=3.63\times10^2\alpha_{\textrm{ss}}^{-\frac{4}{5}}\xi^{2}\bar{\mu}\dot{M_{13}}^{\frac{1}{5}}\Lambda(r)^{
\frac {1}{5}}
\end{equation}
% %
\begin{equation}
B_{\phi,\textrm{shear}}=0.75\gamma \mu_{20}^{-\frac{5}{7}} 
\dot{M_{13}}^{\frac{6}{7}} M_1^{\frac{3} {7}}r^{-3}(1-\omega_s \xi^{-1}r^{3/2}) \rm{T}
\end{equation}
\begin{equation}
B_{\phi,\textrm{dyn}}=12.0\epsilon \alpha_{\textrm{ss}}^{\frac{1}{20}} \xi^{\frac{2}{5}}
\gamma_{\textrm{dyn}}^{\frac{1}{2}}\bar{\mu}^{\frac{3}{16}}\mu_{20}^{-\frac{3}{4}} \dot{M_{13}}^{\frac{4}{5}} 
M_1^{\frac{5} {8}} \Lambda(r)^ {\frac{17}{40}} r^{-\frac{21}{16}} \rm{T}
\end{equation}
\begin{equation}
B_{\textrm{R}}=1.2\times 10^{-5} 
\alpha_{\textrm{ss}}^{\frac{4}{5}}\xi^{-\frac{7}{5}}\bar{\mu}^{-\frac{3}{4}}\mu_{20}^{-\frac{4}{7}} 
\dot{M_{13}}^{\frac{38}{35}} M_1^{-\frac{5} {14}} \Lambda(r)^ {\frac{-7}{10}}\xi^{-1} r^{-\frac{11}{4}} 
\rm{T}.
\end{equation}
It is easy to note that when $\xi=1$ we regain the Keplerian form.  
% % % % % % % % % % % % %
We now present a comparison between quasi keplerian and keplerian structural equation solution.
% % % % % % % % % % % % %
%SD
\begin{figure*}
\gridline{\fig{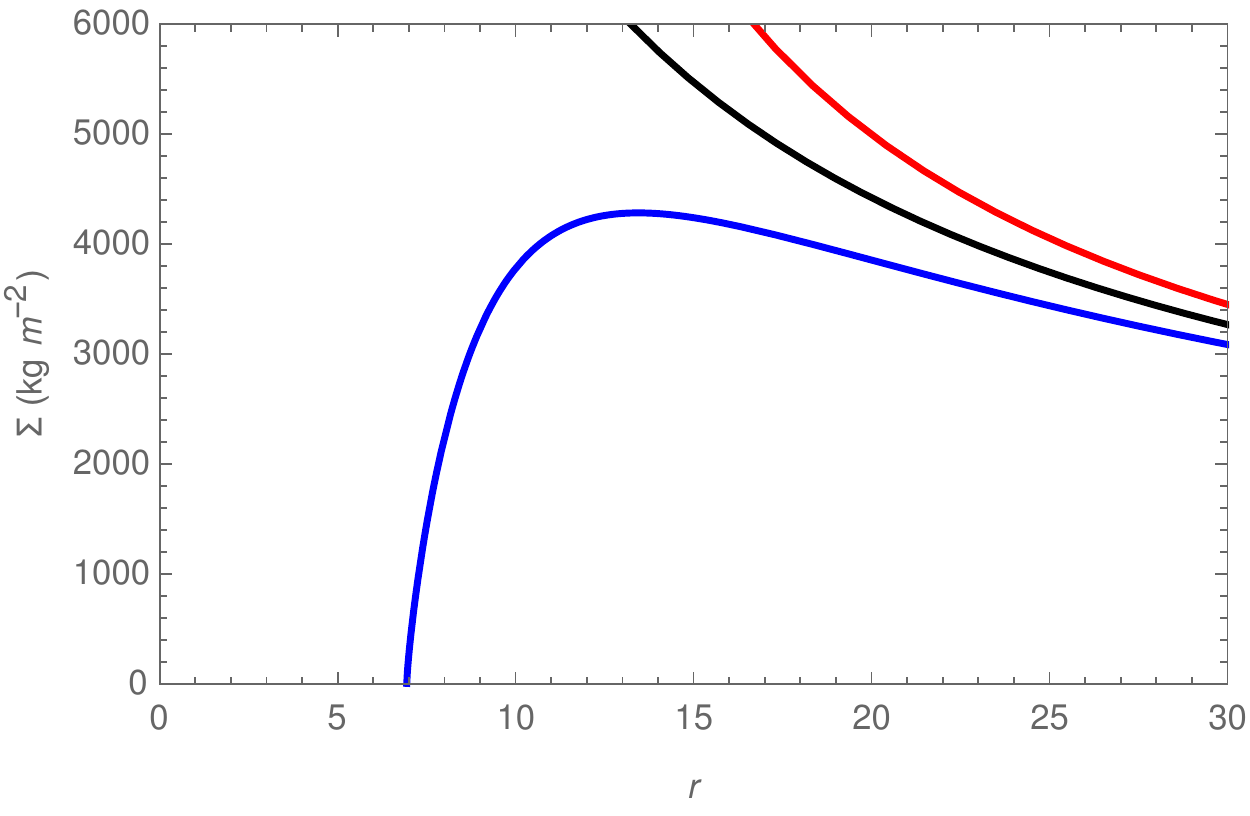}{0.3\textwidth}{(a)$\xi=0.8$}
          \fig{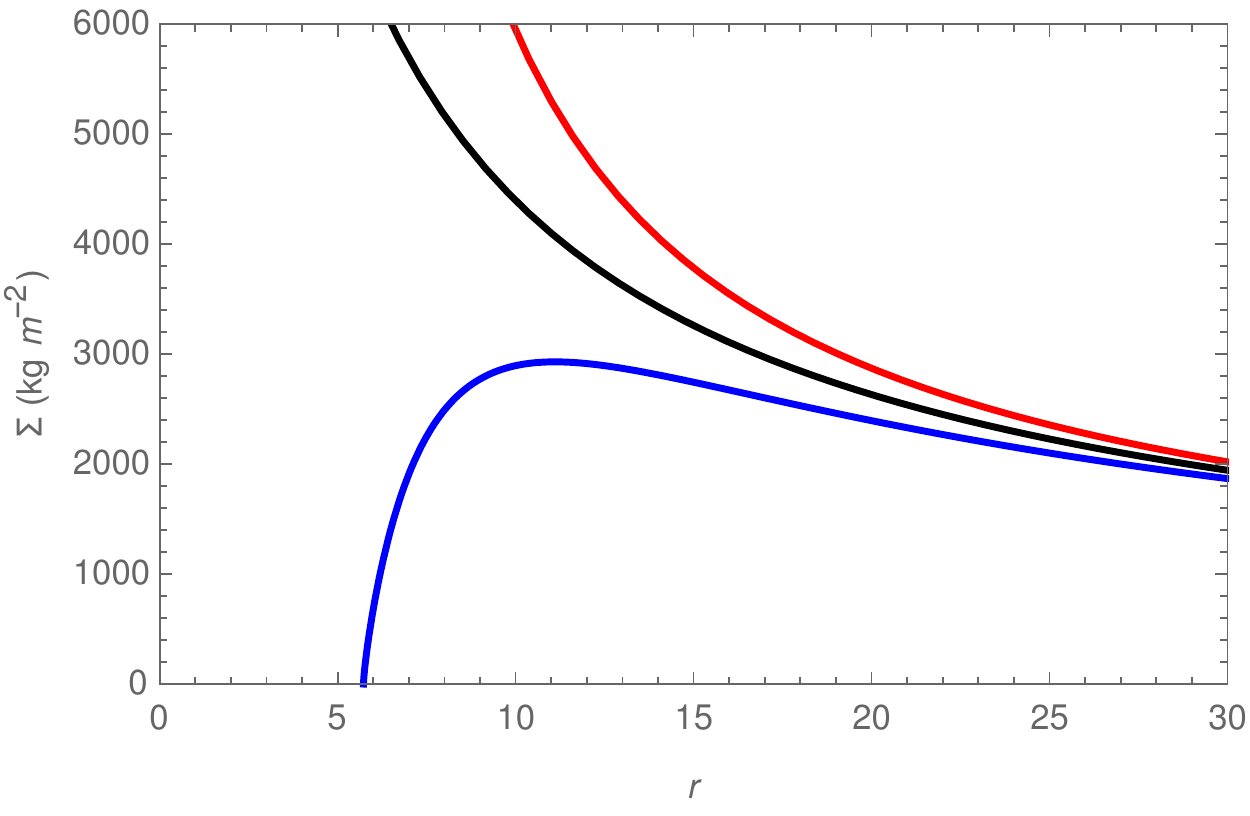}{0.3\textwidth}{(b)$\xi=1.0$}
          \fig{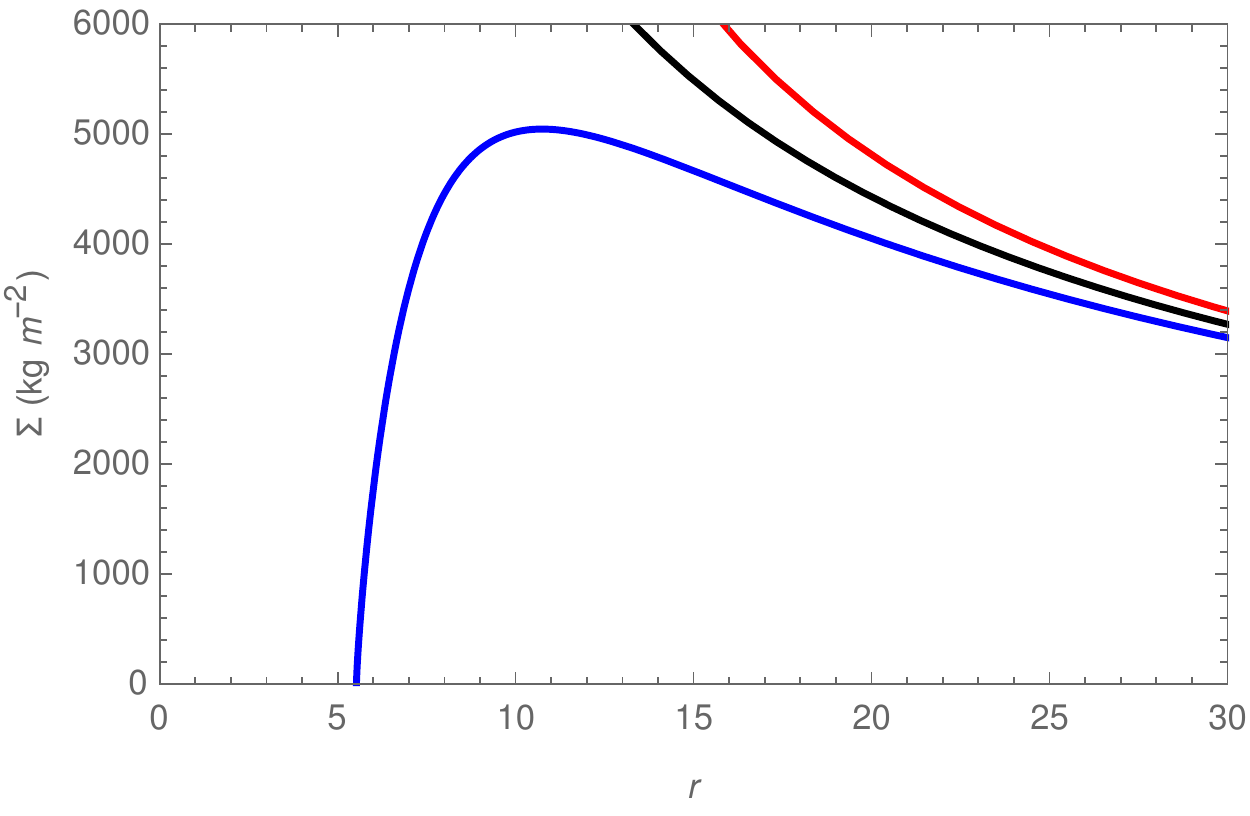}{0.3\textwidth}{(c)$\xi=1.2$}
          }
\gridline{\fig{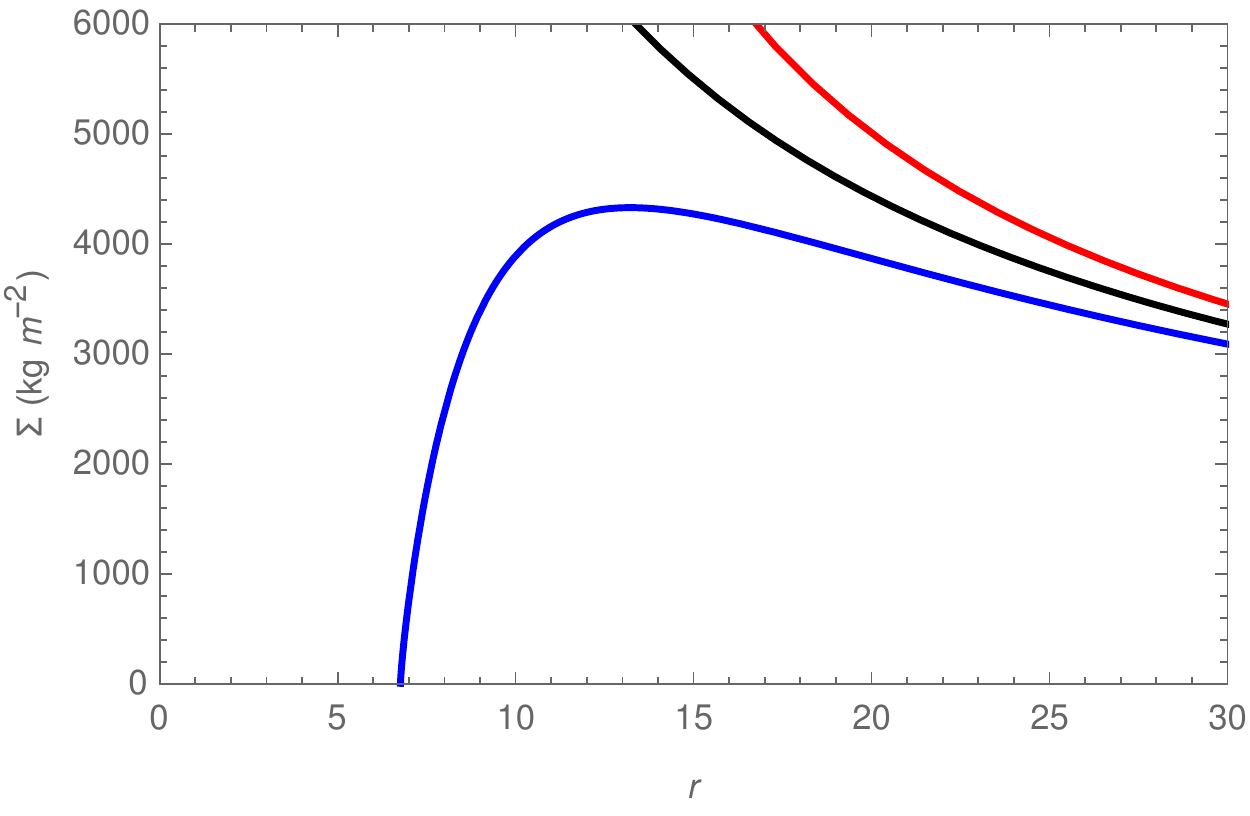}{0.3\textwidth}{(d)$\xi=0.8$}
          \fig{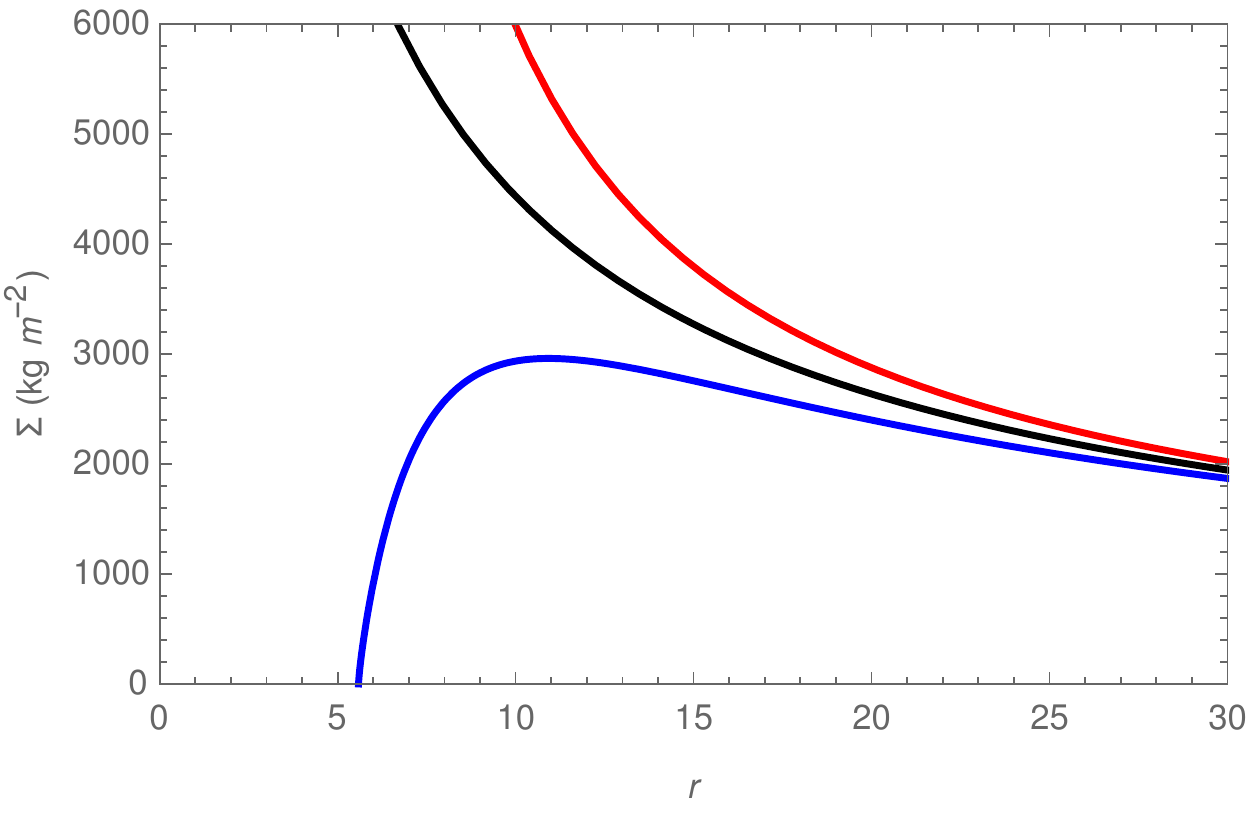}{0.3\textwidth}{(e)$\xi=1.0$}
          \fig{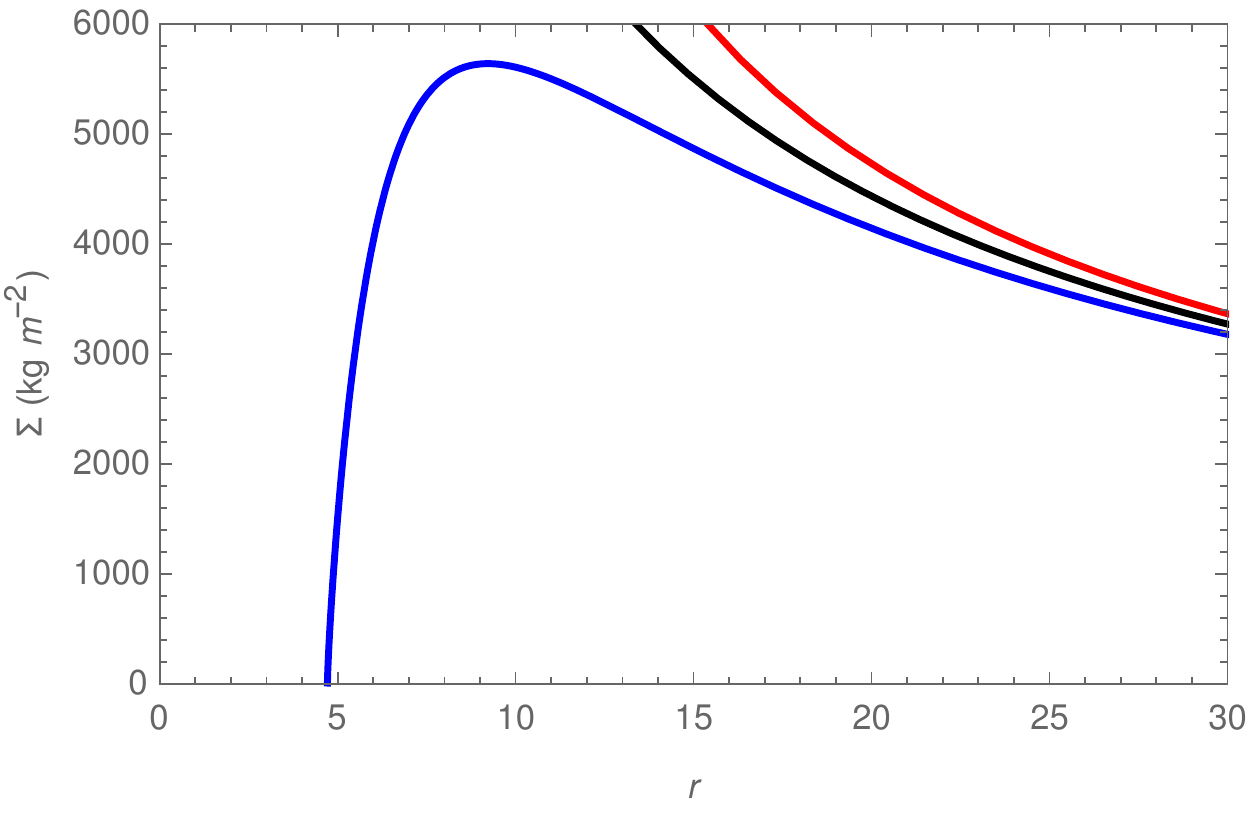}{0.3\textwidth}{(f)$\xi=1.2$}}
\caption{Variation of $\Sigma(r)$ with radial distance for a neutron star with a spin period of: 7s ($top~panel$), 100s 
($bottom~panel$). The magnetic field generated by the dynamo are shown with: $\epsilon=-1.0$ 
blue thick, $\epsilon=0$ black, and $\epsilon=1.0$ red thick}\label{SD}
\end{figure*}
% 
%
%
%Temperature
\begin{figure*}
\gridline{\fig{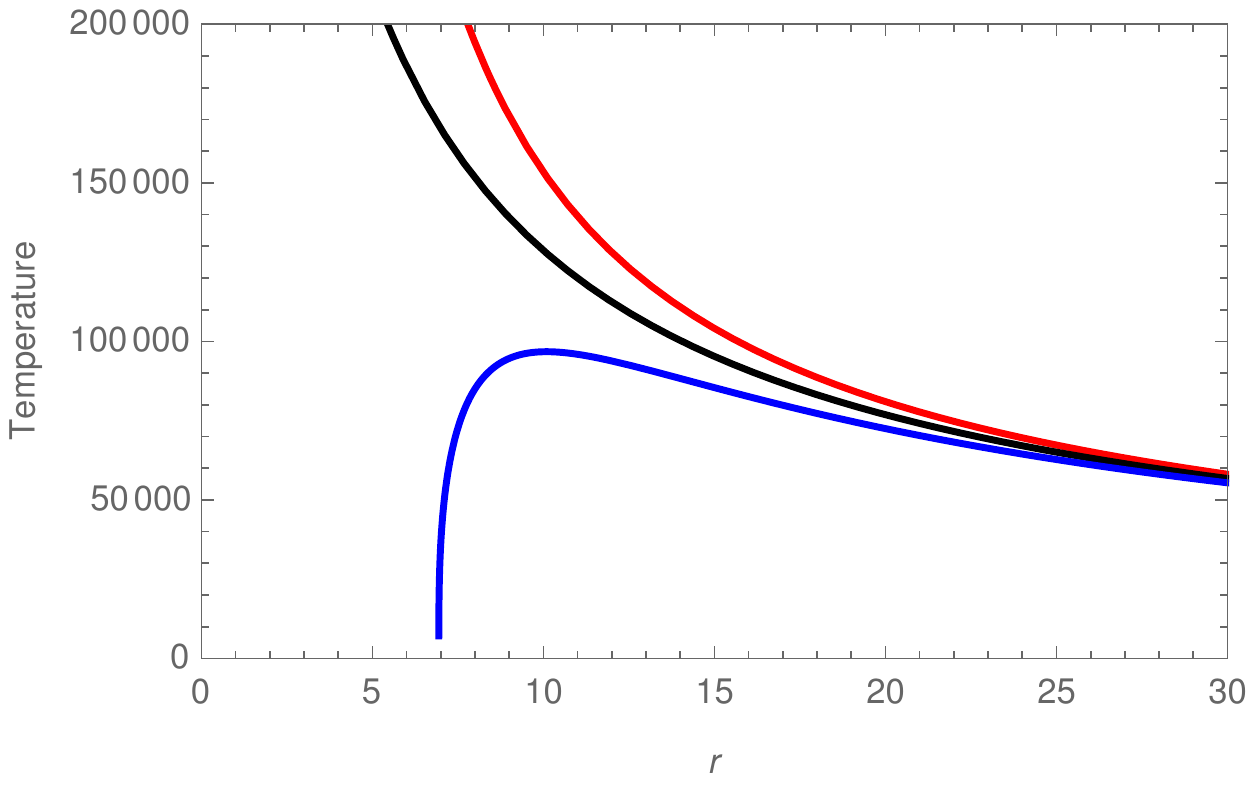}{0.3\textwidth}{(a)$\xi=0.8$}
          \fig{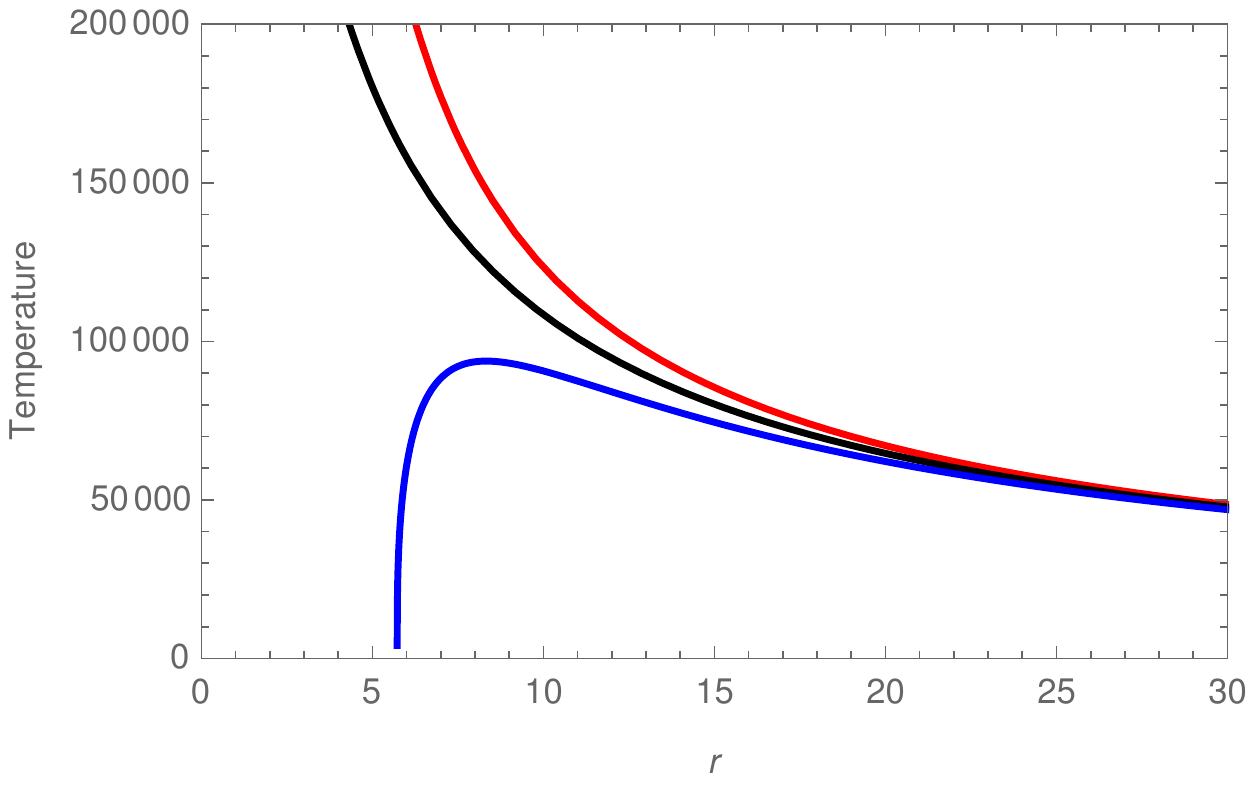}{0.3\textwidth}{(b)$\xi=1.0$}
          \fig{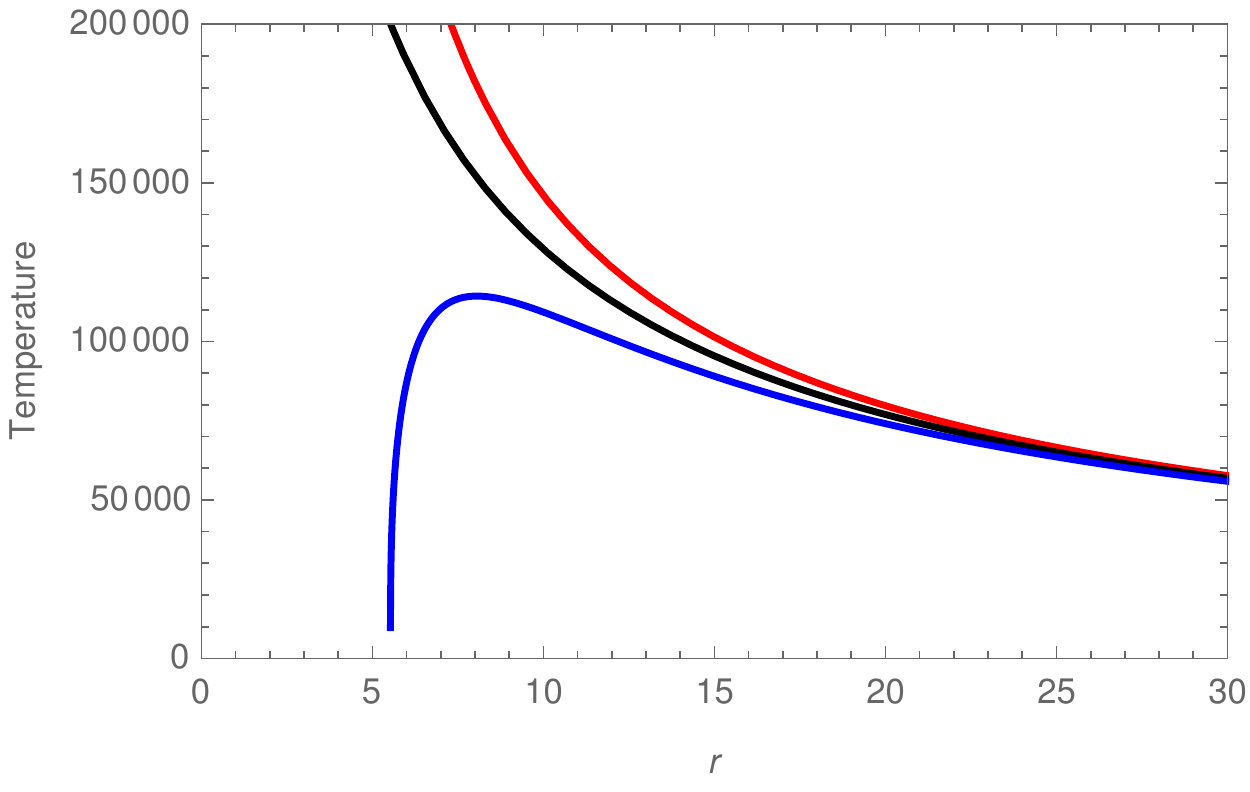}{0.3\textwidth}{(c)$\xi=1.2$}
          }
\gridline{\fig{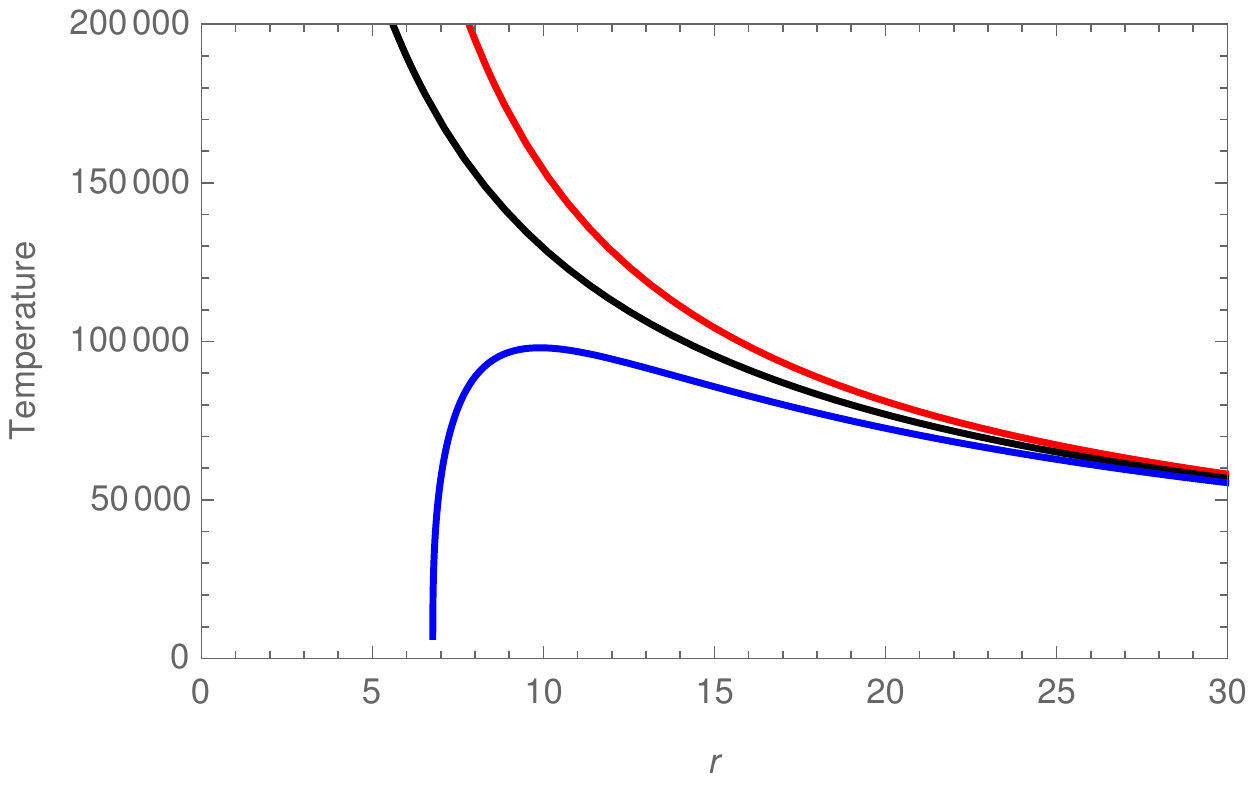}{0.3\textwidth}{(d)$\xi=0.8$}
          \fig{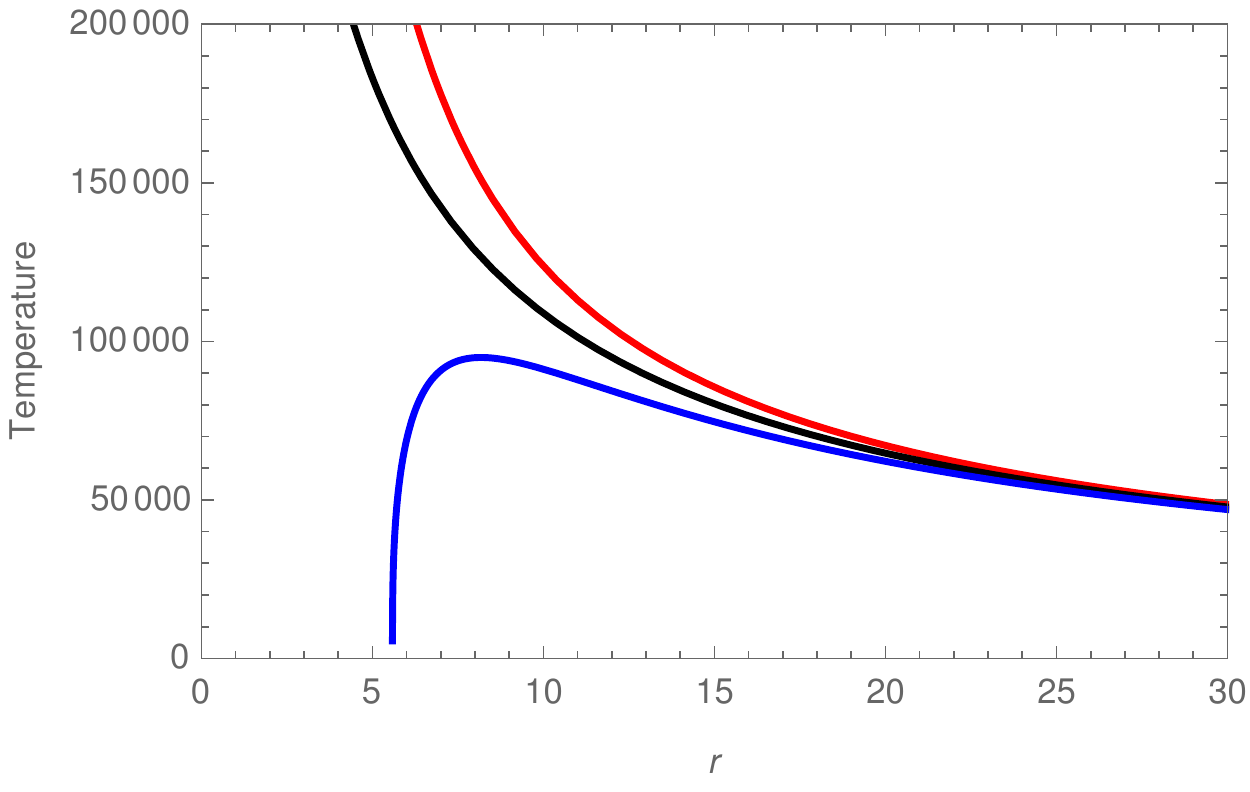}{0.3\textwidth}{(e)$\xi=1.0$}
          \fig{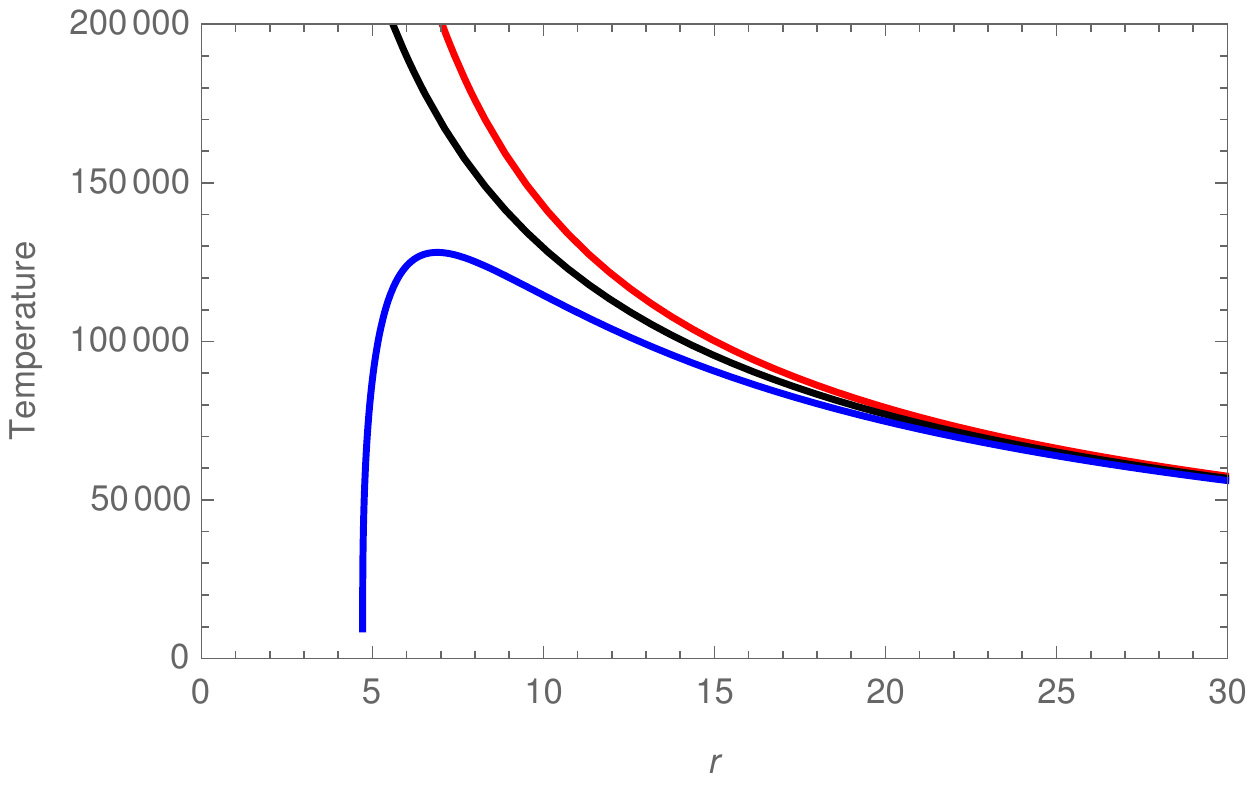}{0.3\textwidth}{(f)$\xi=1.2$}
          }
\caption{Variation of Temperature with radial distance for a neutron star with a spin period of: 7s ($top~panel$), 100s 
($bottom~panel$). The magnetic field generated by the dynamo are shown with: $\epsilon=-1.0$ 
blue thick, $\epsilon=0$ black, and $\epsilon=1.0$ red thick}\label{T}
\end{figure*}

\begin{figure}[h]
\gridline{\fig{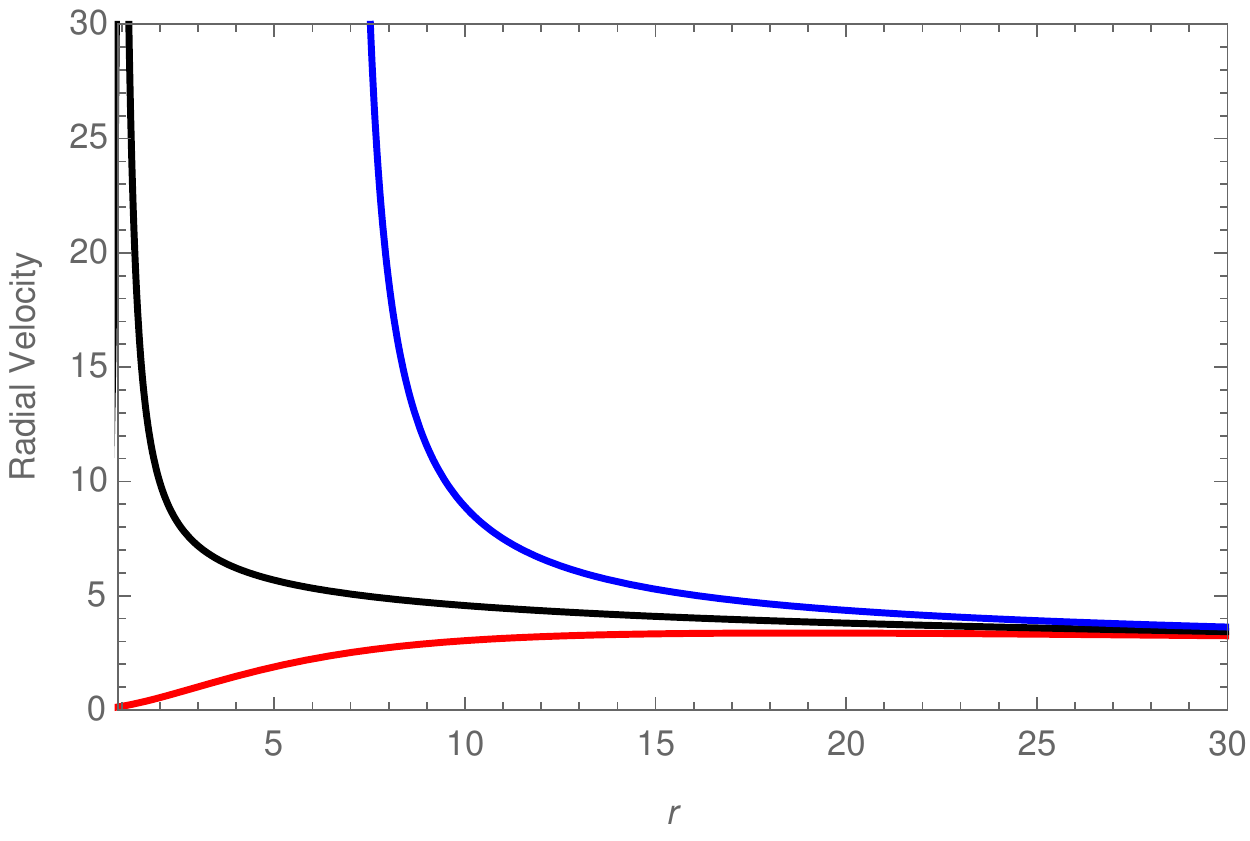}{0.3\textwidth}{(a)$\xi=0.8$}
          \fig{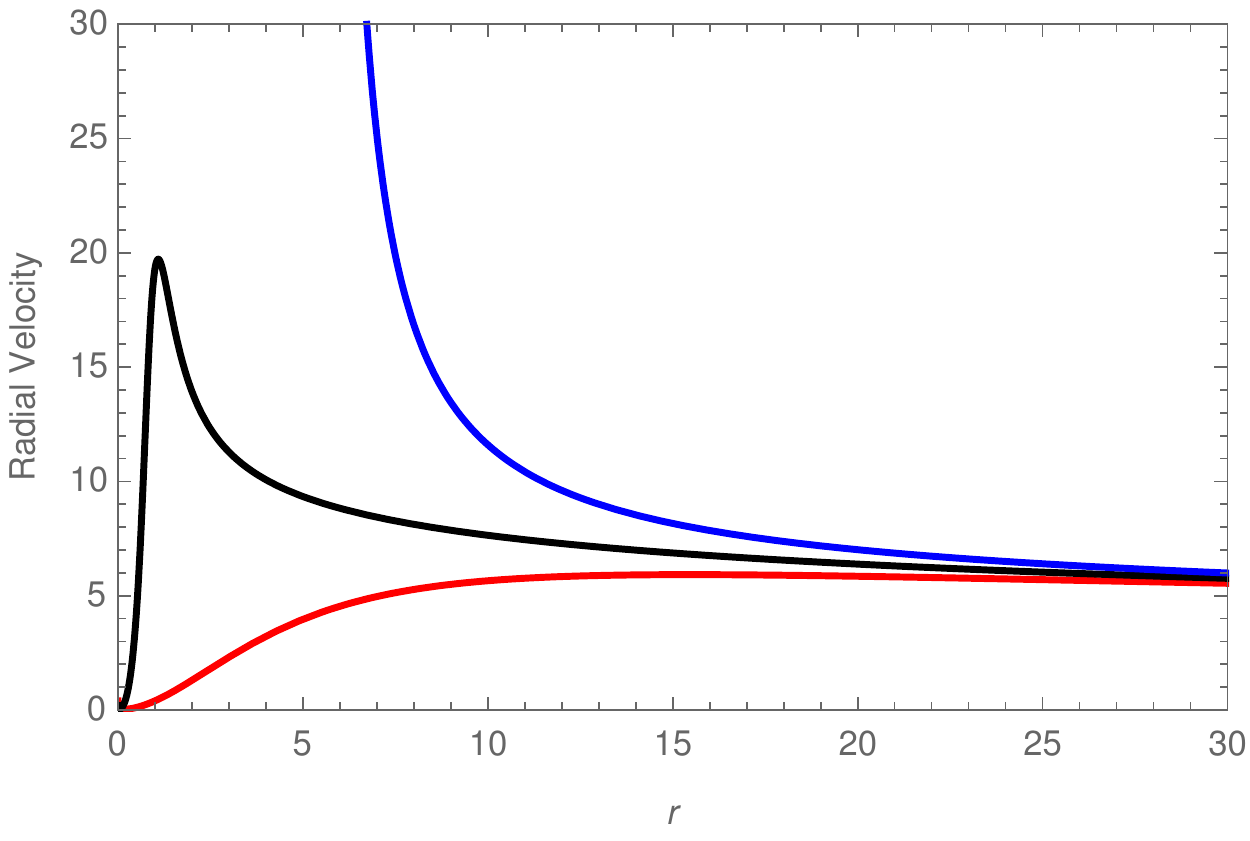}{0.3\textwidth}{(b)$\xi=1.0$}
          \fig{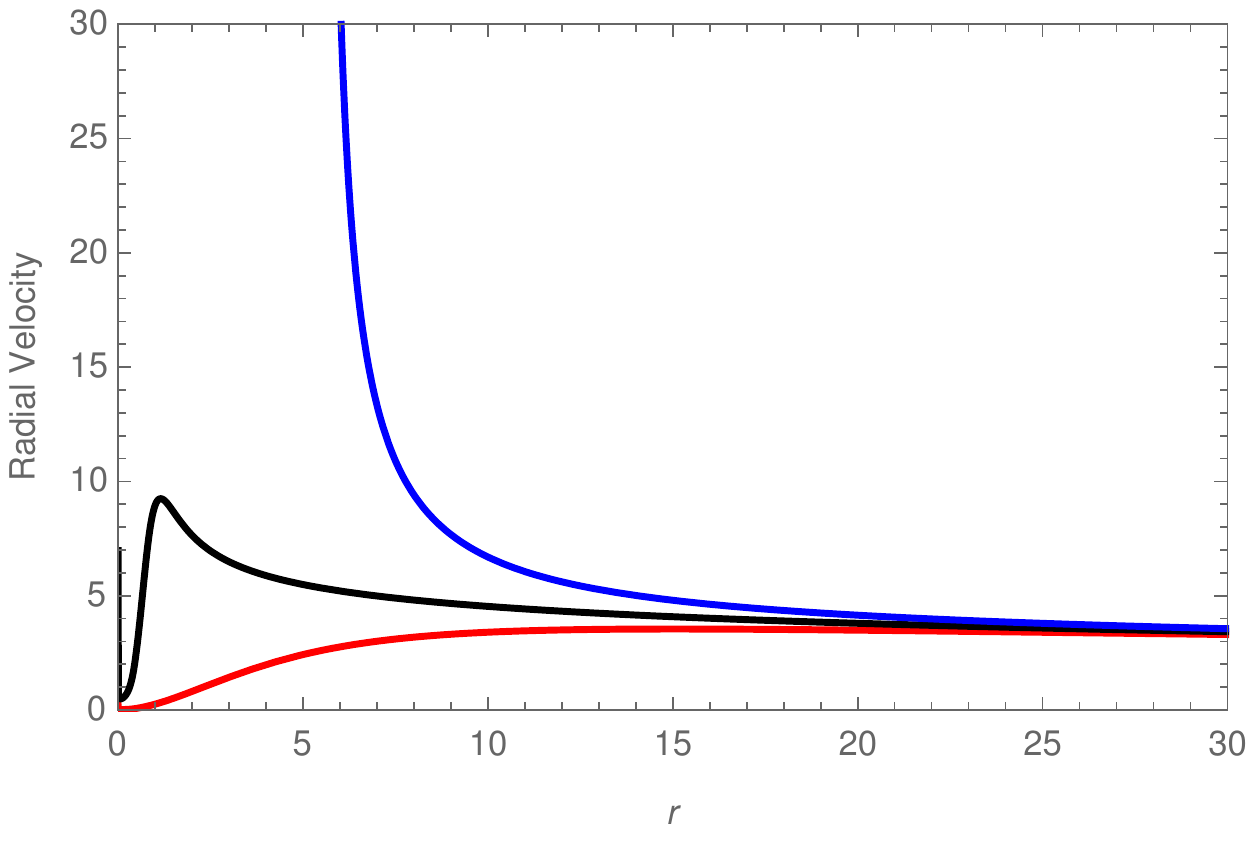}{0.3\textwidth}{(c)$\xi=1.2$}
          }
\gridline{\fig{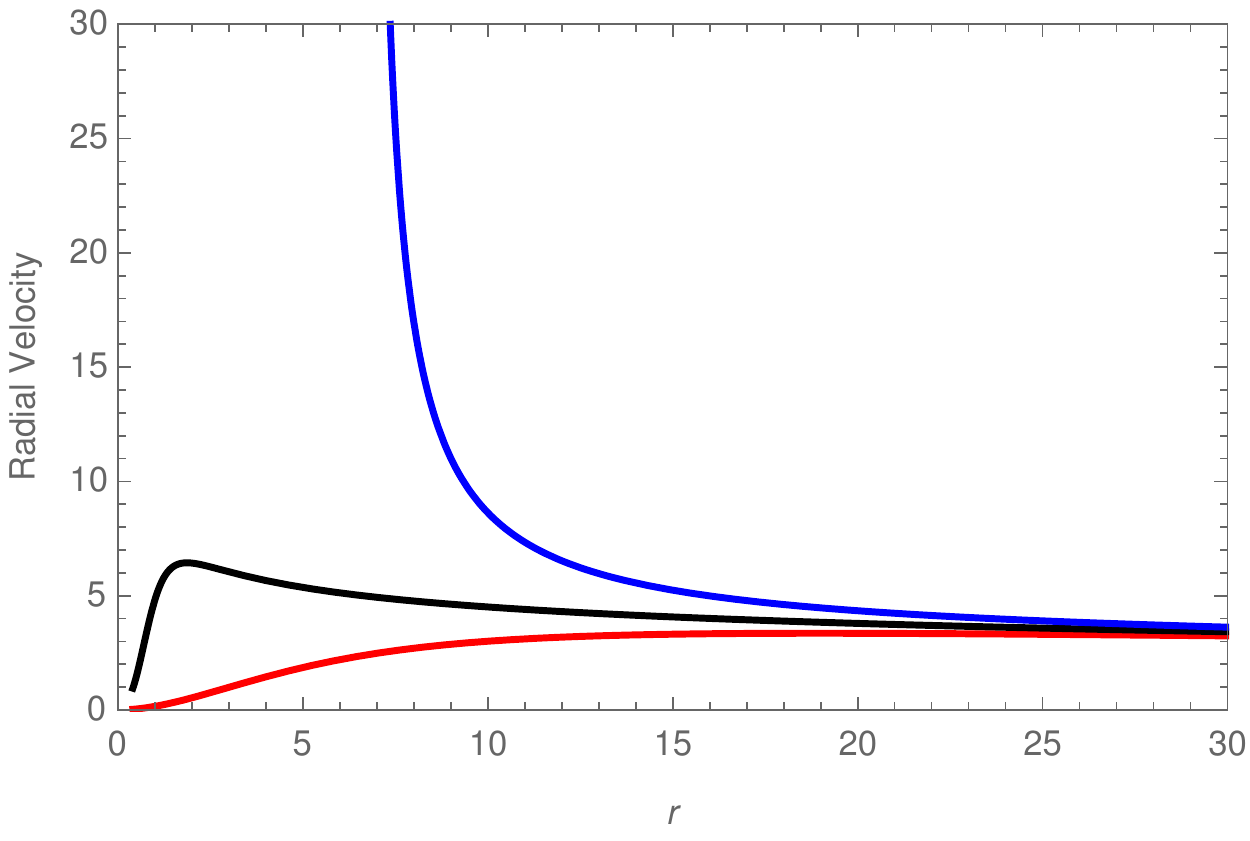}{0.3\textwidth}{(d)$\xi=0.8$}
          \fig{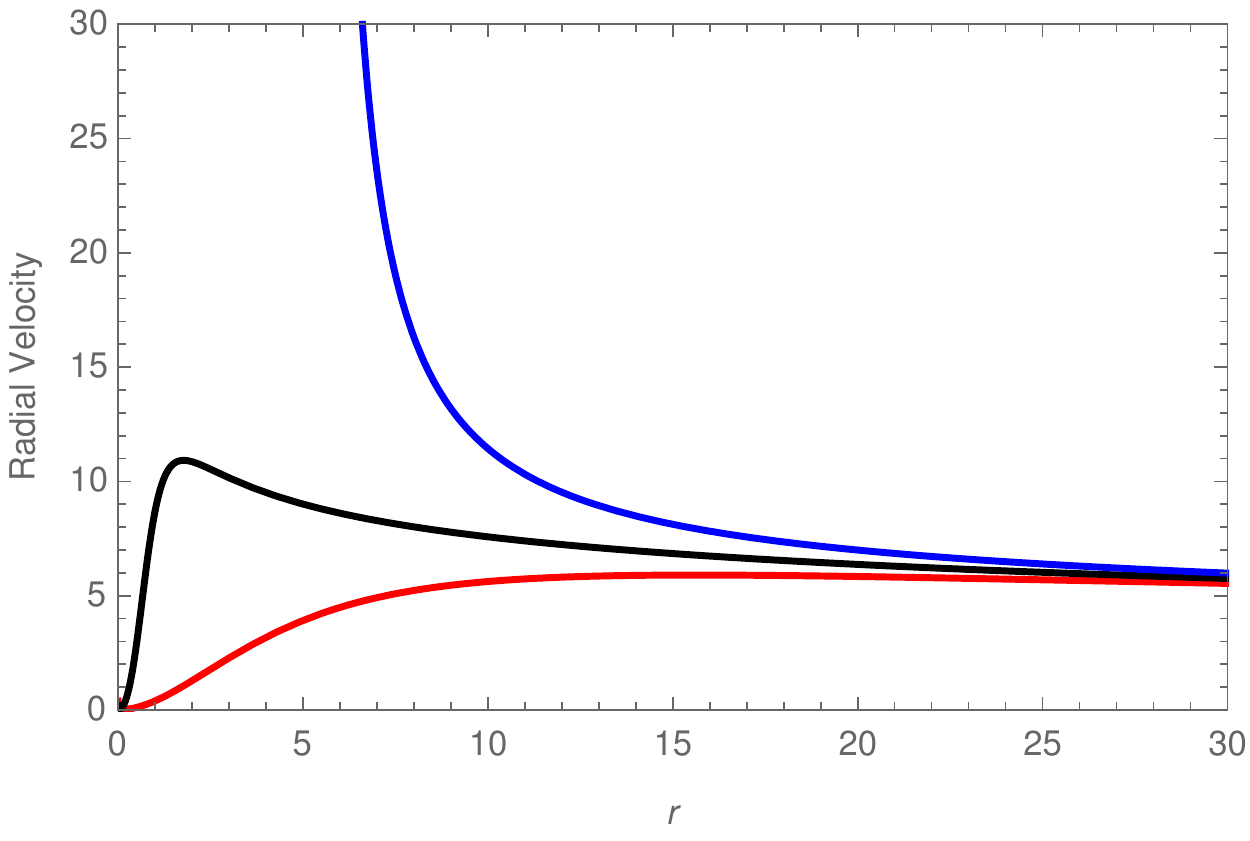}{0.3\textwidth}{(e)$\xi=1.0$}
          \fig{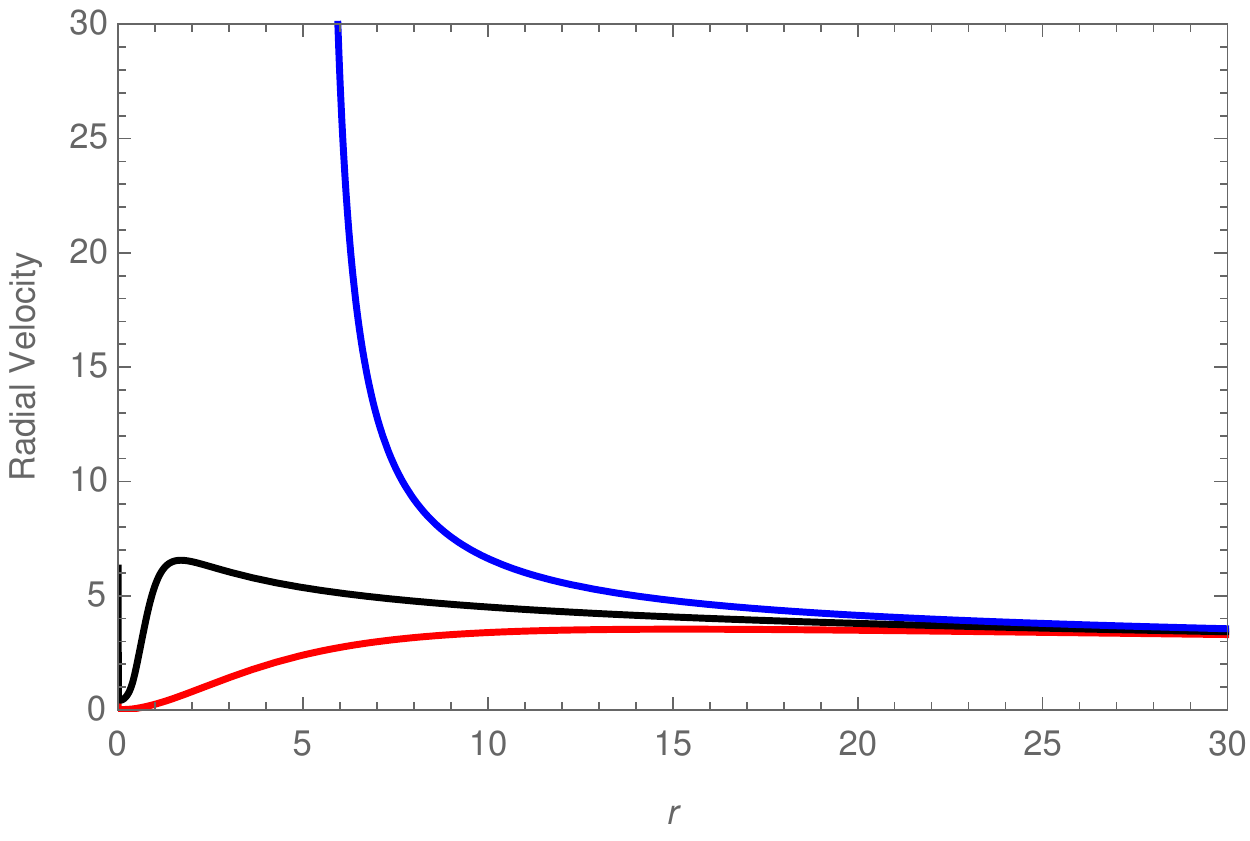}{0.3\textwidth}{(f)$\xi=1.2$}
          }
\caption{Variation of $V_R(r)$ with radial distance for a neutron star with a spin period of: 7s ($top~panel$), 100s 
($bottom~panel$). The magnetic field generated by the dynamo are shown with: $\epsilon=-1.0$ 
blue thick, $\epsilon=0$ black, and $\epsilon=1.0$ red thick}\label{RV}
\end{figure}

In Figure \ref{SD}, we run surface density for spin periods $P=7$ and 100s for changing dynamo parameter 
$\epsilon=-1,0,1$. For $\xi=1.0,$ our results correspond to those of \cite{Tessema2010}. 
Closer to the neutron star, surface density is a purely decreasing function of $r$ for $\epsilon=0~{\rm{and}}~1$. 
With  $\epsilon=-1$, $\Sigma(r)$ develops a local maximum which is observed to increase for $\xi=1.2$ and to decrease for 
$\xi=0.8$. 
The local maximum for either period as the disc deviates from Keplerian motion has no significant change.
The high surface density, $\xi=1.2$ results into a hot flow (Figure \ref{T}) and a corresponding drop in radial velocity 
(Figure \ref{RV}) thus creating pressure gradients.

%
%RV
% 
% 
%Toroidal field
\begin{figure}[ht]
\gridline{\fig{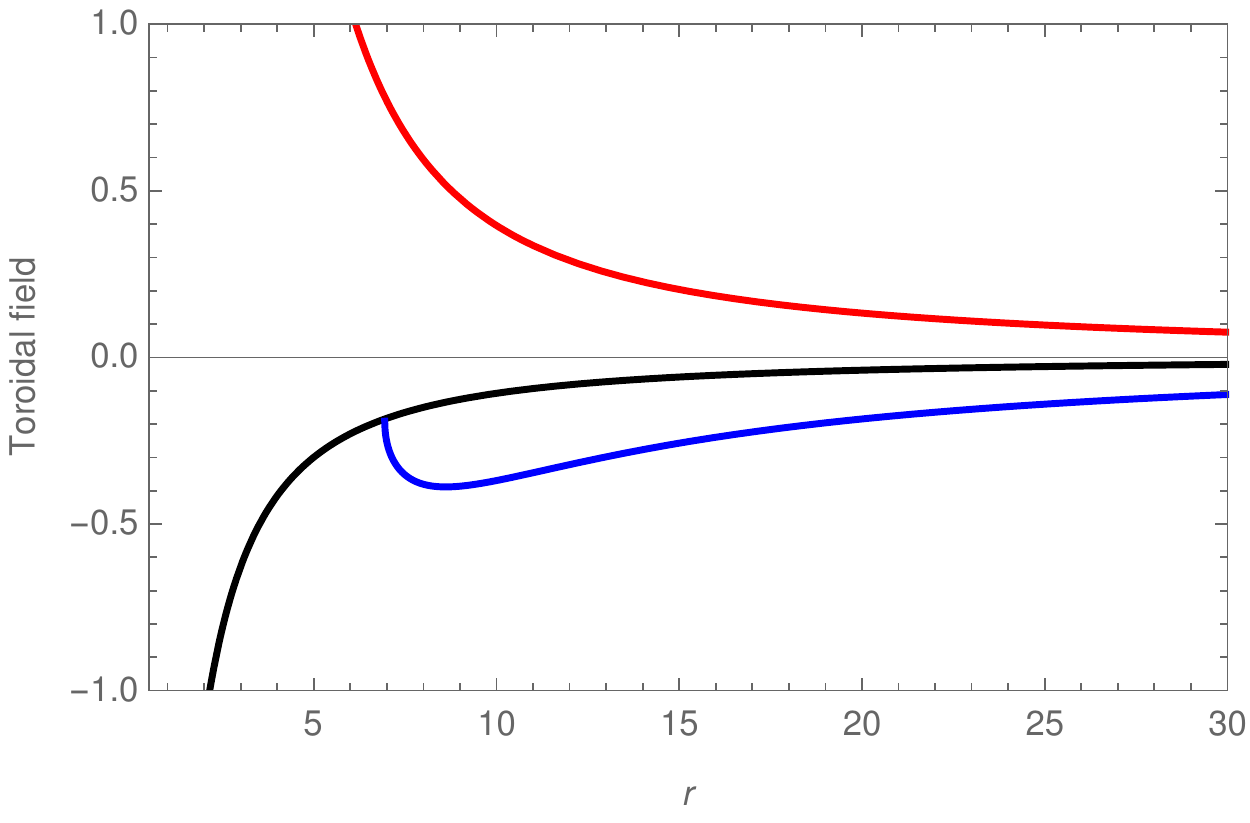}{0.3\textwidth}{(a)$\xi=0.8$}
          \fig{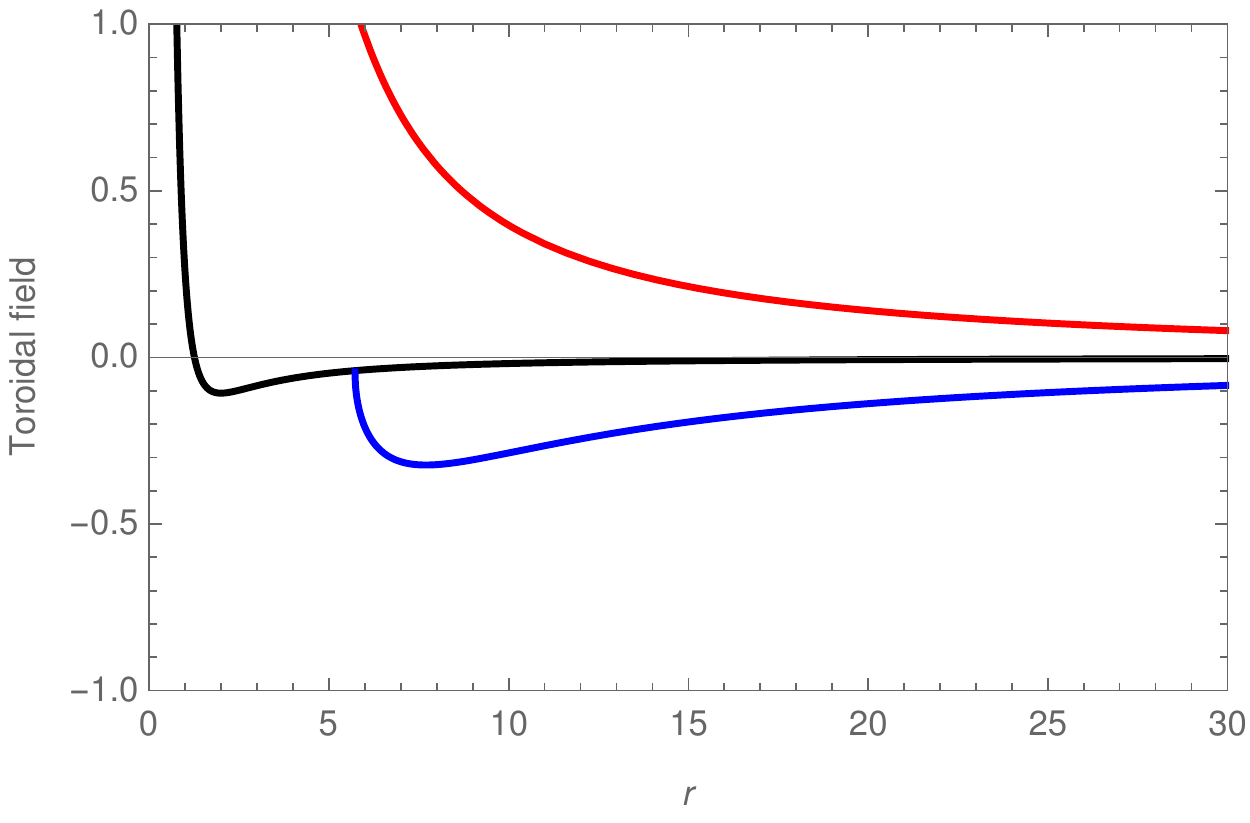}{0.3\textwidth}{(b)$\xi=1.0$}
          \fig{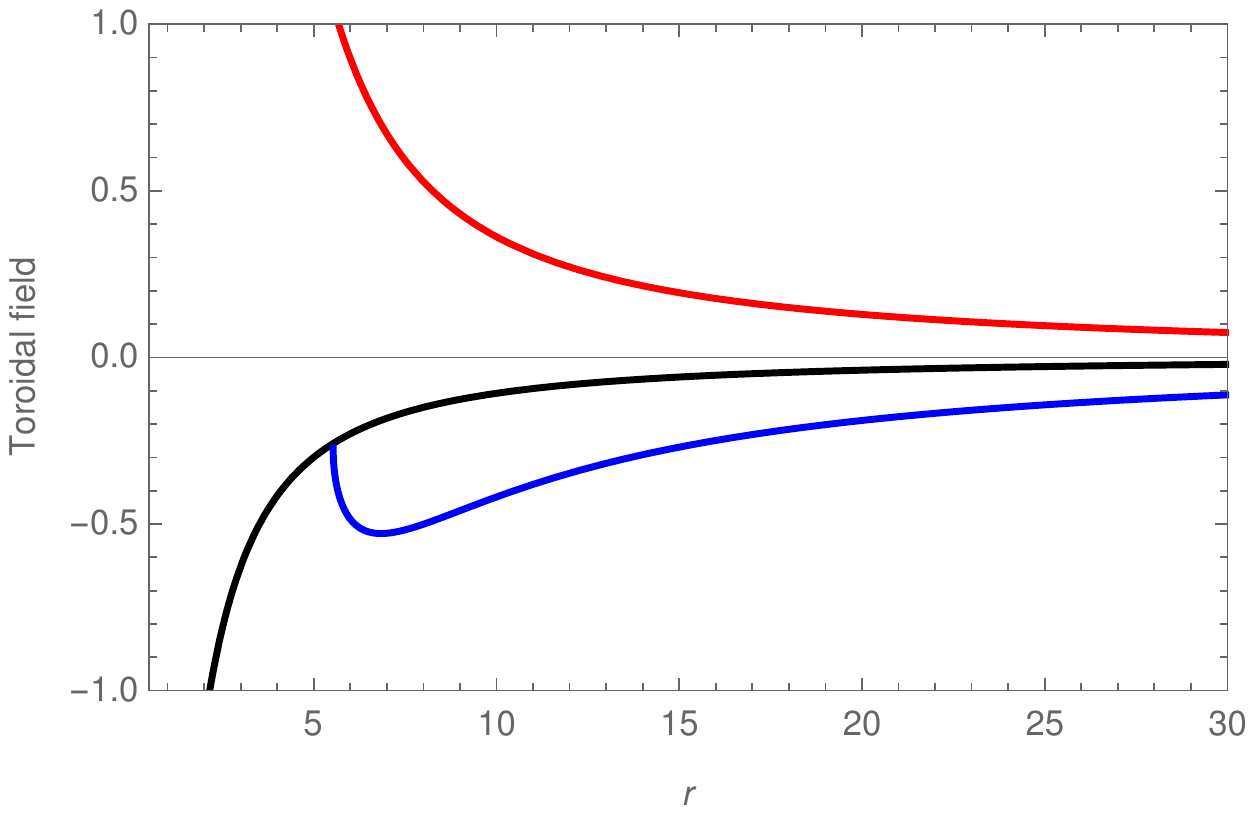}{0.3\textwidth}{(c)$\xi=1.2$}
          }
\gridline{\fig{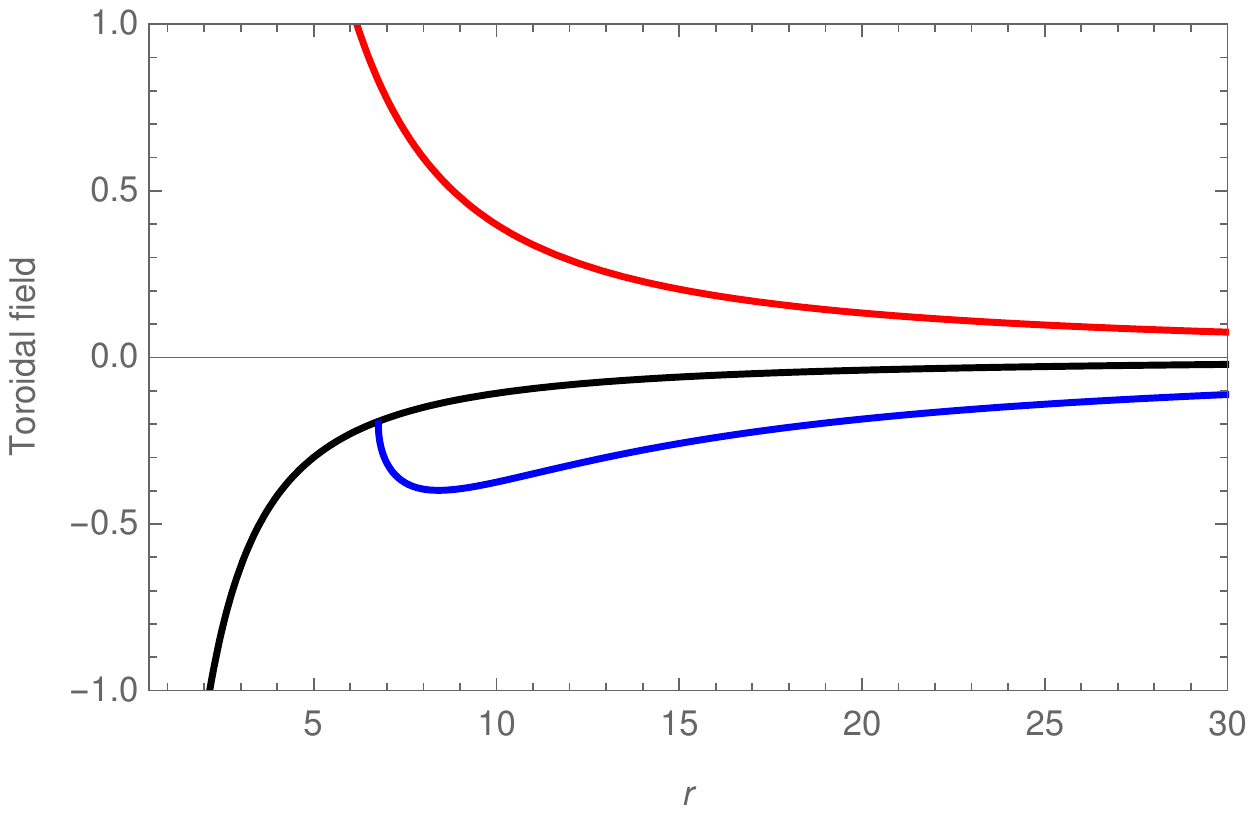}{0.3\textwidth}{(d)$\xi=0.8$}
          \fig{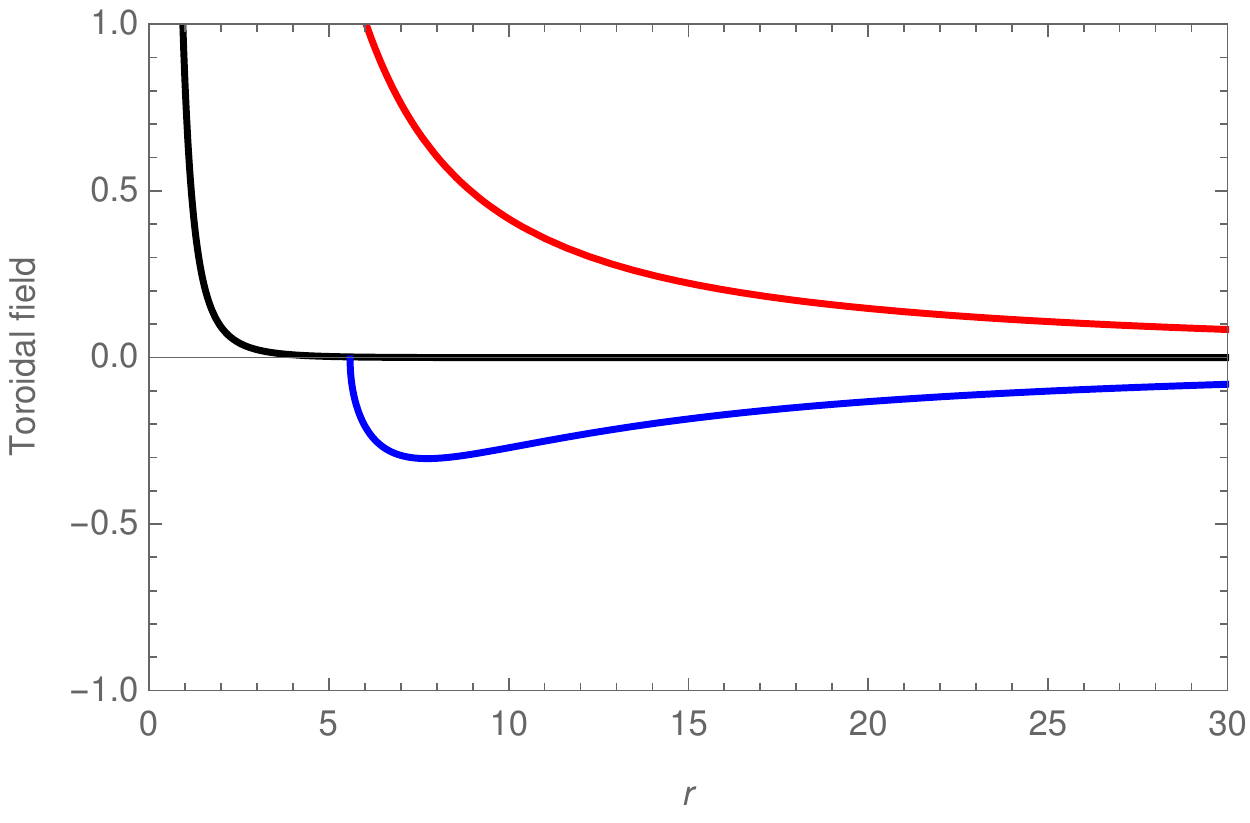}{0.3\textwidth}{(e)$\xi=1.0$}
          \fig{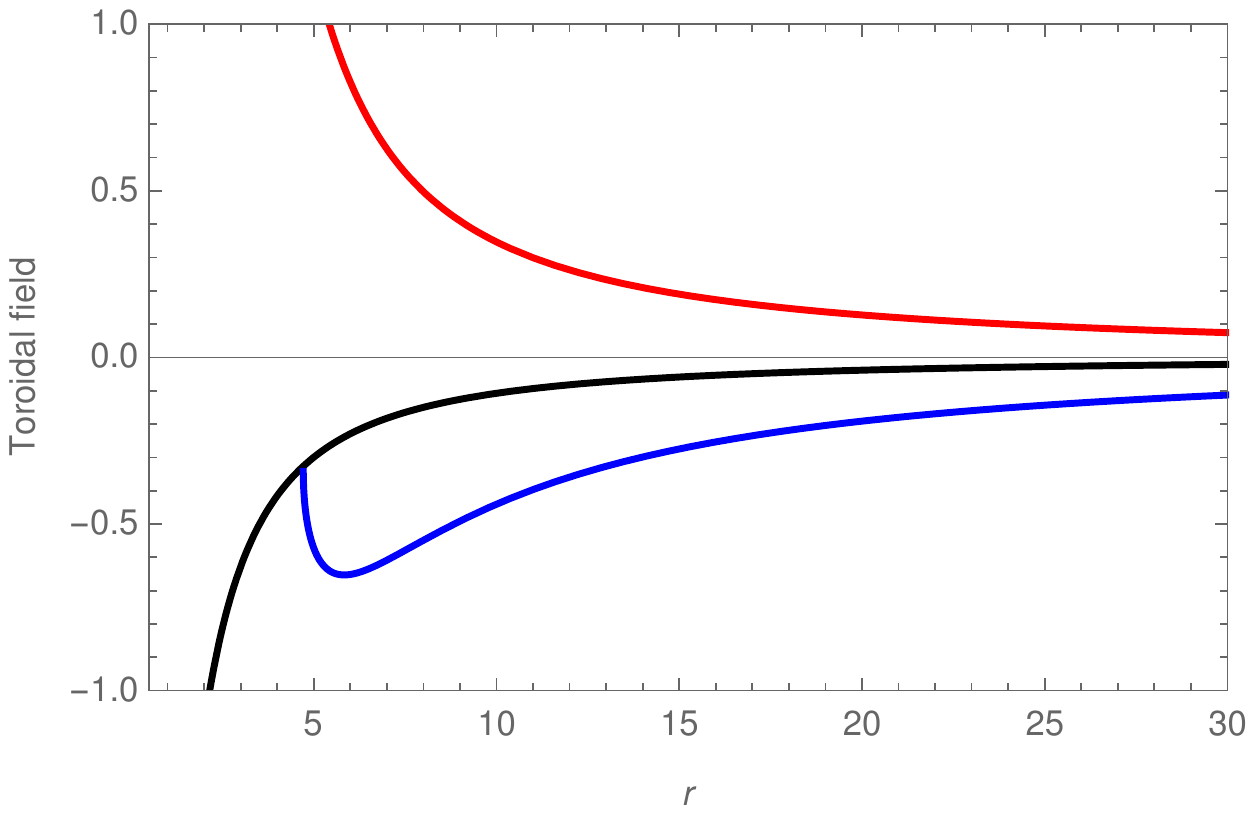}{0.3\textwidth}{(f)$\xi=1.2$}
          }
\caption{Variation of Toroidal field with radial distance for a neutron star with a spin period of: 7s ($top~panel$), 
100s ($bottom~panel$). The magnetic field generated by the dynamo are shown with: $\epsilon=-1.0$ 
blue thick, $\epsilon=0$ black, and $\epsilon=1.0$ red thick}\label{TF}
\end{figure}%

Figure \ref{TF} shows variation of toroidal field with radial distance. Here we note that as the disc transits to 
quasi-Keplerianity, the  magnitude of the toroidal field increases.
Thus, deviation from Keplerian motion has a significant impact on the magnitude of magnetic torques on the 
neutron star. 

\subsection{Effect of pressure gradient force}
In this section we analyze the effect of pressure gradient force for a gas pressure dominated 
case. 
We obtain the pressure gradient equation by vertically integrating Eq. (\ref{b2ra}) to get 
 \begin{equation}
  \frac{\partial \Pi}{\partial R} =-\Sigma\left[v_R\frac{\partial v_R}{\partial R}-\frac{v_\phi^2}{R}\right] 
       - \frac{ \Sigma GM}{R^2} + \left[\frac{B_zB_R}{\mu_0}\right] _{z=-H}^{z=+H},
  \label{b2int}
 \end{equation}
where $\Pi=\int_{-H}^{+H}Pdz$ and $B_R|_{z=-H}=-B_R|_{z=+H}$.  $B_R|_{z=+H}$ means that $B_R$ is evaluated in the upper 
disc plane. The third term on the right hand side represents dominant radial magnetic force.
Using Eq. (\ref{quasi}), (\ref{Br}), (\ref{Bz}) and (\ref{b1mdot})  in Eq. (\ref{b2int}) we obtain
\begin{equation}
% \begin{split}
 \frac{\partial \Pi}{\partial R} 
=\Sigma (\xi^2-1) \frac{ GM}{R^2}+\left(\frac{\dot{M}}{2\pi}\right)^2 \left[\frac{ \Sigma^{-1}}{ R^3} - 
\frac{1}{R^2}\frac{\partial \Sigma^{-1}}{\partial R} \right]
+   
\left(\frac{\dot{M}}{\pi}\right)\left[\frac{\mu^2}{\mu_0}\xi^{-1}\Sigma^{-1}(GM)^{-\frac{1}{2}}R^{-\frac{13}{2}}\right]. 
\label{PG}
% \end{split}
\end{equation}
Transforming and simplifying equation (\ref{PG}) yields: 
 \begin{equation}
%  \begin{split}
   \frac{\partial \Pi}{\partial R} = D_1  (\xi^2-1)  \Lambda(r)^{\frac{7}{10}}  r^{-\frac{11}{4}}
 + D_2  \Lambda(r)^{-\frac{7}{10}}  r^{-\frac{9}{4}}
+D_3\xi^{-1}   \Lambda(r)^{-\frac{7}{10}} r^{-\frac{23}{4}},\label{RPGn}
% \end{split}
 \end{equation}
where
$D_1=\left[\frac{3}{2}\left(\frac{243\kappa_0}{512\sigma}\right)^{-\frac{1}{10}}   
\alpha_{\textrm{ss}}^{-\frac{4}{5}}\left(\frac{k_B}{\bar{\mu}m_p}\right)^{-\frac{3}{4}}(GM)^{\frac{5}{4}}\dot{M}^{\frac{
7} {10}}R_A^{ -\frac{11}{4}} \right ]$, 
 \\
$D_2=\left[\frac{1}{4\pi^2}\left(\frac{243\kappa_0}{512\sigma}\right)^{\frac{1}{10}}   
\alpha_{\textrm{ss}}^{\frac{4}{5}}\left(\frac{k_B}{\bar{\mu}m_p}\right)^{\frac{3}{4}}(GM)^{-\frac{1}{4}}\dot{M}^{-\frac{13
} {10}}R_A^{ -\frac{9}{4}} \right ]$, and \\
$D_3=\left[\frac{2}{3\pi}\frac{\mu^2}{\mu_0}\left(\frac{243\kappa_0}{512\sigma}\right)^{\frac{1}{10}}   
\alpha_{\textrm{ss}}^{\frac{4}{5}}\left(\frac{k_B}{\bar{\mu}m_p}\right)^{\frac{3}{4}}(GM)^{-\frac{3}{4}}
\dot{M}^{\frac{3}{10}}R_A^{ -\frac{23}{4}}\right ]$.
 % % % 

Equation (\ref{RPGn}) is the pressure gradient equation for this model. 
The first term on the RHS is the dominant term resulting from the difference between Keplerian and 
quasi-Keplerian angular momentum. We also note here that in a Keplerian state the PGF will vanish. This is in agreement 
with the definition of a thin Keplerian accretion disc \citep{Campbell1992,Frank2002}.
Although pressure is a scalar quantity, PGF is a vector normal to the local disc and is directed along the disc plane.
% % % % % % 

\begin{figure*}[ht]
\gridline{\fig{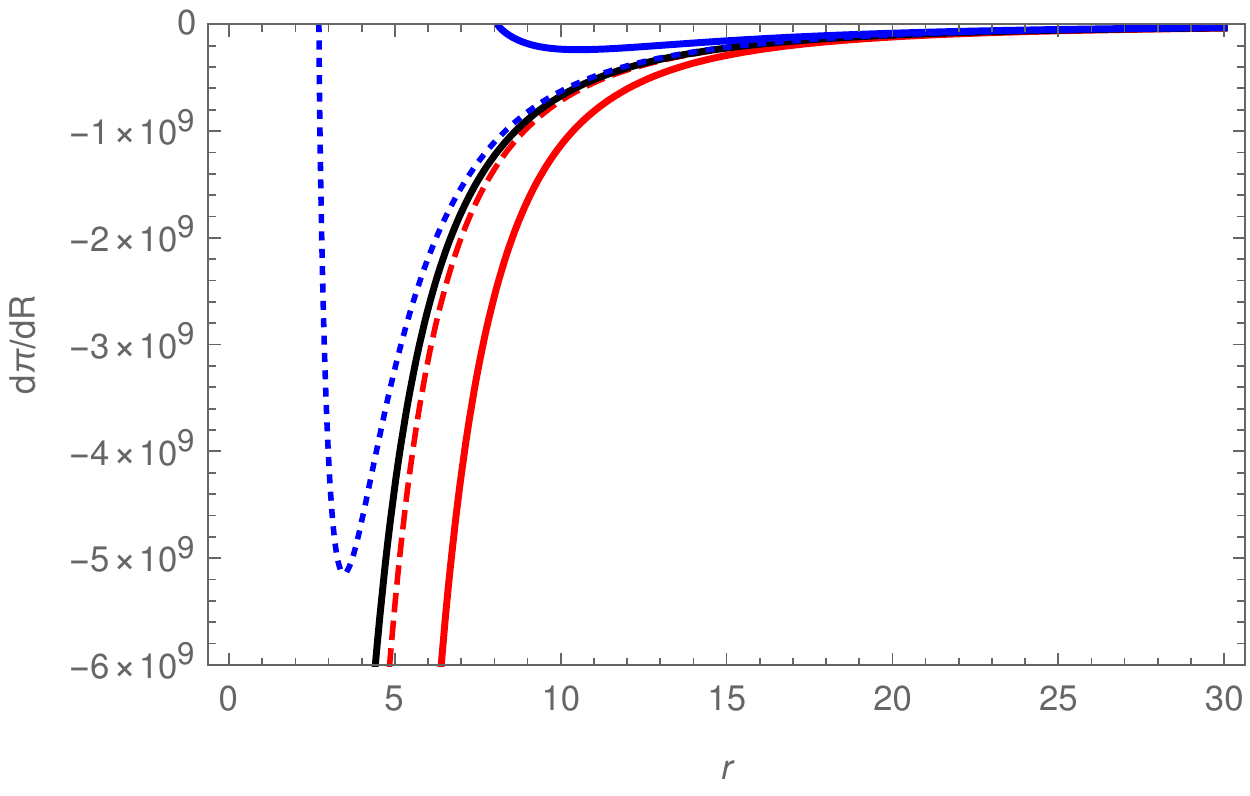}{0.3\textwidth}{(a)$\xi=0.8$}
         \fig{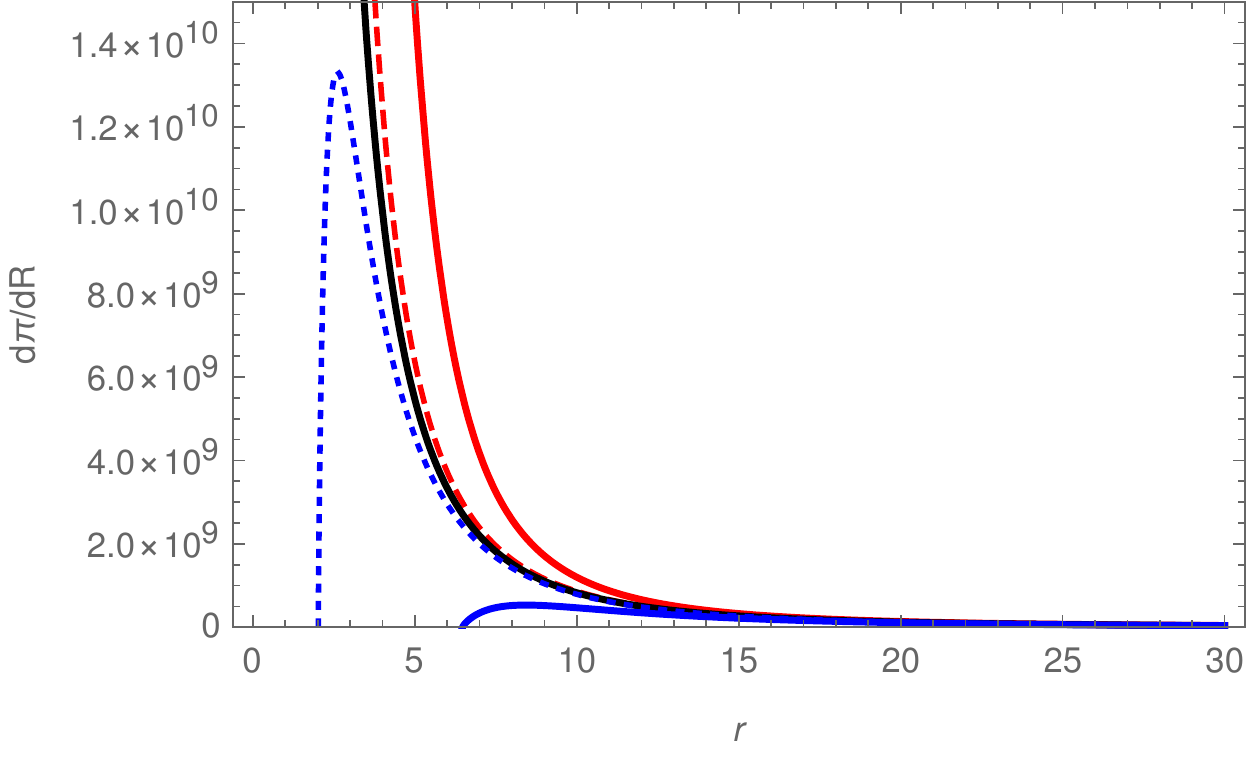}{0.3\textwidth}{(b)$\xi=1.2$}
          }
\gridline{\fig{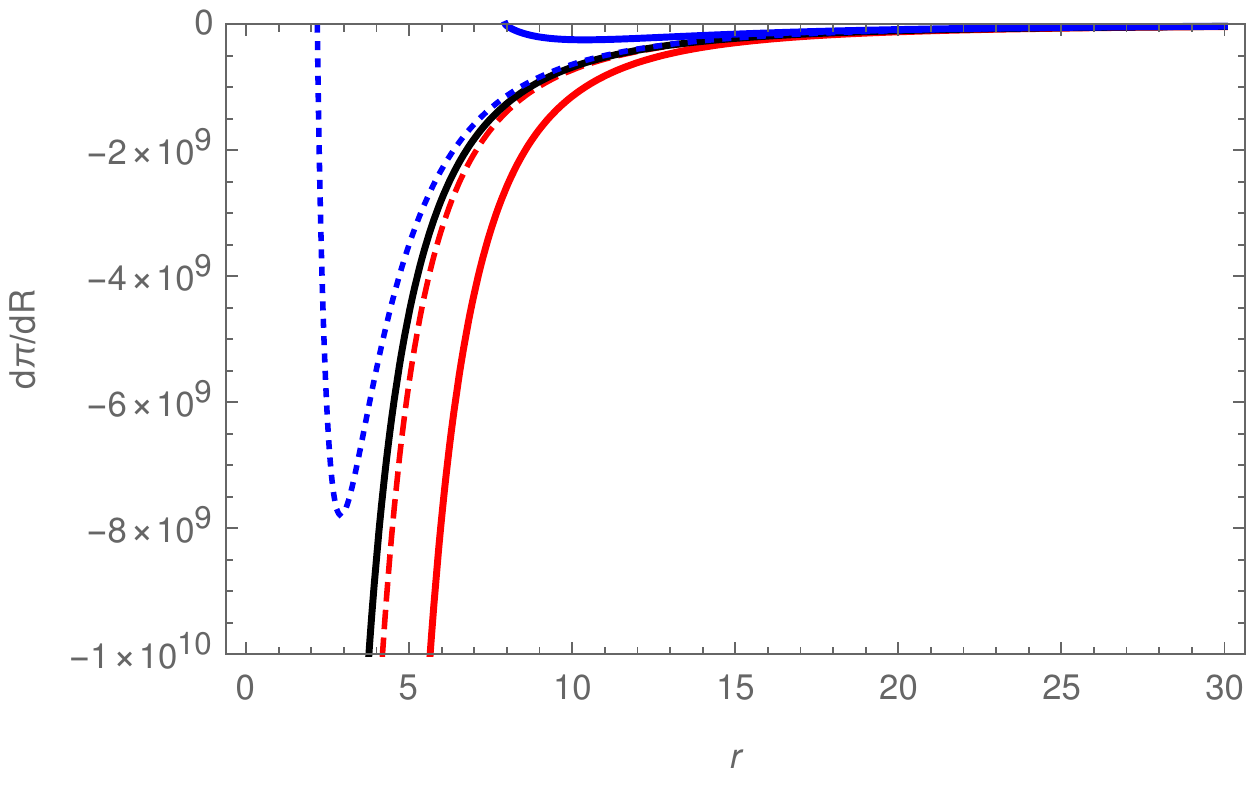}{0.3\textwidth}{(c)$\xi=0.8$}
        \fig{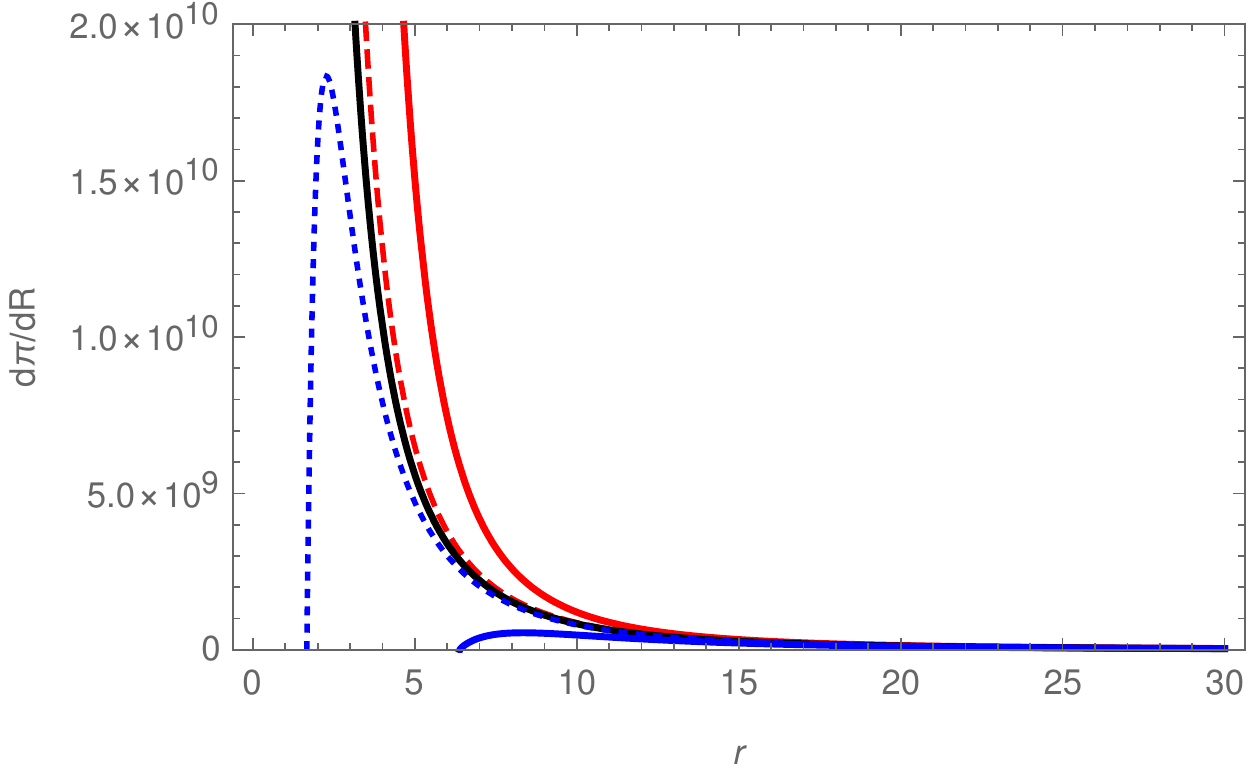}{0.3\textwidth}{(d)$\xi=1.2$}
          }
\caption{Pressure gradient force ($\partial \Pi/\partial R$) as a function of radial distance for a neutron star with a 
spin period of: 7.0s ({\it{top panel}} (a) \& (b)), 100s ({\it{bottom panel}} (c) \& (d)). The magnetic field generated 
by the dynamo are shown with: $\epsilon=-1.0$ 
blue thick, 
$\epsilon=-0.1$ blue dotted, $\epsilon=0$ black, $\epsilon=0.1$ red dashed and $\epsilon=1.0$ red thick}\label{pgf7s}
\end{figure*}

We plot the PGF in Figure (\ref{pgf7s} for $P$=7s and for $P$=100s) with varying values 
of $\epsilon ~{\rm{and}}~\xi$. Our results show that for $\xi=0.8$ the PGF is negative and becomes positive for 
$\xi=1.2$.
PGF increases with period and the local maximum ($\xi=1.2$), minimum ($\xi=0.8$) decreases with increasing radial 
distance. 
This reversal translates into torque reversal as the disc makes a transition to and from Keplerian motion. 
A quasi-Keplerian motion can show observed dynamical scenarios. We shall have a detailed discussion in the next section 
after finding the net torque acting on the neutron star.

As the disc switches between Keplerianity; $\xi=1.2$ (increased azimuthal velocity) and $\xi=0.8$ (reduced azimuthal 
velocity), pressure differences force matter from areas of high pressure to the areas of low pressure see 
(Figure \ref{T}). 
% % % % % % % % 
Therefore, in addition to 
magnetic torques, pressure gradient force can contribute to the total torque exerted on the neutron star. 
% PGF
% %
\subsection{Torques on a neutron star in a quasi-Keplerian disc}
The torques on a neutron star range from magnetic torques to material torques. These are obtained from  Eq. 
(\ref{azimuthal}) and (\ref{b2int}), by multiplying by $2\pi R$ and then vertically integrate from $R_{in}$ to 
$R_{out}$.
The torque contribution from the pressure gradient force is:
\begin{equation}
% \begin{split}
  N_{PGF}=-2\pi\int_{R_{in}}^{R_{out}} \Sigma\left[v_R\frac{\partial v_R}{\partial R}-\frac{v_\phi^2}{R}\right]R dR  
- 2\pi \int_{R_{in}}^{R_{out}}\frac{ \Sigma GM}{R} dR+ 2\pi\int_{R_{in}}^{R_{out}} 
\left[\frac{B_zB_R}{\mu_0}\right] _{z=-H}^{z=+H}RdR,
% \end{split}
 \label{npgf}
 \end{equation}
 where $N_{PGF}=\int_{R_{in}}^{R_{out}}2\pi R \left( \frac{\partial \Pi}{\partial R}\right)dR$, while the other torques 
are:

\begin{equation}
% \begin{split}
\left[-\xi \dot{M}\sqrt{GMR}\right]_{R_{in}}^{R_{out}}=
\int_{R_{in}}^{R_{out}} \frac{4\pi}{\mu_0} [B_z (B_{\phi, \rm{dyn}}+B_{\phi, \rm{shear}})]R^2dR
-\left[ 3\pi \xi(\nu \Sigma) (GMR)^\frac{1}{2}\right]_{R_{in}}^{R_{out}},
% \end{split}
\label{torque}
\end{equation}
where $R_{in}$ is the position of the disc inner edge.
% % % % % % %  
The LHS of equation (\ref{torque}) shows the rate at which angular momentum is transported past the inner and outer edge 
of the accretion disc while the first and second term on the RHS represents the effect of magnetic and viscous stresses 
respectively. Considering only exchange of angular momentum between the neutron star and accretion disc, the torques are 
obtained as:
\begin{equation}
N_{adv}(r_i)=2.6\times 10^{26}\xi\mu_{20}^{2/7}M_1^{3/7}\dot{M}_{13}^{6/7}r_i^{1/2}
\end{equation}
\begin{equation}
 N_{shear}=-7.5\times 10^{26}  \gamma \mu_{20}^{\frac{2}{7}} M_1^{\frac{3}{7}}\dot{M}_{13}^{\frac{6}{7}} 
\int_{r_{i}}^{\infty} \left\{\frac{1-(\omega_s/ \xi) r^{\frac{3}{2}}}{r^{4}}\right\}dr
\end{equation}
\begin{equation}
 N_{dyn}=1.2\times 10^{28}  \epsilon \gamma_{\rm{dyn}}^{\frac{1}{2}} \alpha_{\textrm{ss}}^{\frac{1}{20}} 
\bar{\mu}^{\frac{3}{16}}\mu_{20}^{\frac{1}{4}} M_1^{\frac{5}{8}}\dot{M}_{13}^{\frac{4}{5}}\chi^{-\frac{3}{16}} 
\int_{r_{i}}^{\infty} \Lambda^{\frac{17}{40}}r^{-\frac{37}{16}}dr
\end{equation}
\begin{equation}
N_{vis}(r_i)=-2.4\times 10^{27}\xi\mu_{20}^{2/7}M_1^{3/7}\dot{M}_{13}^{6/7}\Lambda(r) r_i^{1/2}.
\end{equation}
% % 
As the disc deviation from Keplerian motion increases, the magnitude of both shear, $N_{shear}$ and dynamo, $N_{dyn}$ 
induced torques increase. $N_{shear}$ changes sign whenever $\xi>\omega_s$ and the contribution from 
this torque vanishes at a point when $\xi=\omega_s$. The viscous and advective torques are reduced for $\xi<1$ and 
amplified for $\xi>1$ by 20\% below and above the Keplerian case respectively. 
The overall effect is that in the non-Keplerian case, the neutron star experiences torques of greater magnitude than it 
is 
for the case of Keplerian.
% 
% % % % % % % % % % % % % % % % % % % 
 
 \subsection{Assessment of torque}
 % % % % % % % % % % % % % % % % % 
 
The total torque exerted on the neutron star, $N_T$ can be expressed in terms of the inner edge position $R_{in}$ as:
\begin{equation}
% \begin{split}
 N_T(R_{in})= N_{PGF}+2.6 \times 10^{26}\xi r_{in}^{\frac{1}{2}}- 2.5\times10^{27} \xi\Lambda_{(r_{in})} 
r_{in}^\frac{1}{2} 
 -3.3\times 10^{46} 
\frac{1}{r_{in}^{3}}\left[1-\frac{2}{\xi}\left(\frac{r_{in}}{r_{co}}\right)^{\frac{3}{2}}\right] 
+ 5.2\times 10^{42}\epsilon \Lambda_{(r_{in})}^{\frac{17}{40}}r_{in}^{-\frac{21}{16}}.
%  \end{split}
\end{equation}
Here $\Lambda_{R_{in}}$ means evaluation at $R_{in}$.
The range between $R_{in}$ to $\infty$ covers both spin up and spin down contributions from magnetic stresses.  
Specifically, $R_{in} \rightarrow R_{co}$ results in a spin-up torque while $R_{co} \rightarrow\infty$ contributes a 
spin-down torque to the neutron star \citep{Wang1987,Wang1995}. 
 
Numerical solutions for torques arising due to interaction of the neutron star and accretion disc are calculated and 
tabulated in Table \ref{table}.

% % % % % % % % % % % 
\begin{table*}[ht]
\begin{minipage}{1.0\textwidth}
\caption{Net torque on a neutron star evaluated at $R_{in}$}
\begin{center}
\footnotesize
\begin{tabular}{@{}lccccccccccr@{}}
\hline\hline
$P_{\rm{spin}}$  & $\xi$ & $\epsilon$ & Case & $R_{in}$ &$N_{adv}$ & $N_{visc}$ & $N_{dyn}$& $N_{shear}$ & 
$N_{PGF}$&  
$N_{\rm{Total}}$\\
\hline\hline
7.0 
& 0.8&1.0 & V&10RA& 6.6 $\times 10^{26}$&-1.2$\times 10^{27}$& 1.2$\times 10^{27}$&-1.7$\times 10^{25}$&-7.0$\times 
10^{23}$&6.4$\times 10^{25}$\\
& & 0.1&V&2.5RA&3.3 $\times 10^{26}$ &-6.2$\times 10^{26}$&7.5$\times 10^{26}$&-1.5$\times 10^{26}$&-5.6$\times 
10^{24}$&3.0$\times 10^{26}$\\
&  & 0 & V&1.0RA& 2.1 $\times 10^{26}$ &-2.0$\times 10^{25}$&0&-3.1$\times 10^{26}$&-2.2$\times 10^{25}$&-1.4$\times 
10^{26}$\\
&  & -0.1 &D&2.5RA&3.3 $\times 10^{26}$&0 &-7.5$\times 10^{26}$&-1.3$\times 10^{26}$&-1.4$\times 10^{25}$&-5.6$\times 
10^{26}$\\
&  & -1.0 & D&7.0RA&5.5 $\times 10^{26}$&0& -1.9$\times 10^{27}$& -2.9$\times 10^{25}$&-6.4$\times 10^{24}$&-1.4$\times 
10^{27}$\\\hline
   % % % % % % % % % % % % % % % % % % % % 
 & 1.0 & 1.0 & V&8.0RA&  7.4 $\times 10^{26}$&-1.4$\times 10^{27}$&1.4$\times 10^{27}$&-1.9$\times 
10^{25}$&0&7.2$\times 10^{26}$\\
& & 0.1 &V&2.0RA&3.7 $\times 10^{26}$ &-7.0$\times 10^{26}$&8.9$\times 10^{26}$&-1.3$\times 10^{26}$&0&4.3$\times 
10^{26}$\\
& & 0 & V&1.0RA&2.6 $\times 10^{26}$ &-1.5$\times 10^{26}$&0&-2.0$\times 10^{26}$&0&-9.0$\times 10^{25}$\\
 &  & -0.1 &D&2.0RA&3.7 $\times 10^{26}$&0&-8.9$\times 10^{26}$&-1.3$\times 10^{26}$&0&-6.5$\times 10^{26}$\\ 
 &   & -1.0 & D&5.7RA& 6.2 $\times 10^{26}$&0&-2.2$\times 10^{27}$&-3.1$\times 10^{25}$&0&-1.6$\times 10^{27}$\\\hline
     % % % % % % % % % % % % % % % % % % % %
& 1.2& 1.0 & V&7.5RA& 8.6 $\times 10^{26}$&-1.6$\times 10^{27}$&1.8$\times 10^{27}$&-1.8$\times 10^{25}$&1.3$\times 
10^{24}$&1.0$\times 10^{27}$\\
& & 0.1 &V&1.9RA& 4.3 $\times 10^{26}$&-8.1$\times 10^{26}$&1.1$\times 10^{27}$&-1.1$\times 10^{26}$&1.0$\times 
10^{25}$&6.2$\times 10^{26}$\\
& &0 &V&1.0RA& 3.1 $\times 10^{26}$&-2.4$\times 10^{26}$&0&-1.3$\times 10^{26}$&2.7$\times 10^{25}$&-3.3$\times 
10^{25}$\\
& & -0.1&D&1.8RA& 4.2 $\times 10^{26}$&0&-1.2$\times 10^{27}$&-1.1$\times 10^{26}$&3.1$\times 10^{25}$&-8.6$\times 
10^{26}$\\
&  & -1.0 &D&5.5RA& 7.4 $\times 10^{26}$&0&-2.7$\times 10^{27}$&-2.7$\times 10^{25}$&1.3$\times 10^{25}$&-2.0$\times 
10^{27}$\\\hline\hline
% % % % % % % % % % % %  
 100 &  0.8&1.0 &V&9.5RA& 6.5 $\times 10^{26}$ &-1.2$\times 10^{27}$&1.3$\times 10^{27}$&-1.0$\times 
10^{24}$&-7.5$\times 10^{23}$&7.5$\times 10^{26}$\\
&   & 0.1 &V&3.8RA&4.1 $\times 10^{26}$&-7.7$\times 10^{26}$&4.4$\times 10^{27}$&-7.8$\times 10^{23}$&-3.0$\times 
10^{24}$&4.0$\times 10^{27}$\\
 &  & 0 &V&1.0RA&2.1 $\times 10^{26}$ &-3.9$\times 10^{26}$&0&2.1$\times 10^{25}$& -2.2$\times 10^{25}$&-1.8$\times 
10^{26}$\\
  & & -0.1 &D&1.8RA&2.8 $\times 10^{26}$ &0&-1.2$\times 10^{27}$&2.6$\times 10^{25}$& -9.1$\times 10^{24}$&-9.0$\times 
10^{26}$\\
 &  & -1.0 &D&7.0RA&5.5 $\times 10^{26}$ &0&-1.9$\times 10^{27}$&-1.4$\times 10^{24}$&-6.2$\times 10^{24}$&-1.4$\times 
10^{27}$\\\hline
     % % % % % % % % % % % % % % % % % % % %
  & 1.0 &1.0&V&7.5RA&  7.2 $\times 10^{26}$&-1.3$\times 10^{27}$&1.6$\times 10^{27}$&-9.3$\times 
10^{23}$&0&1.0$\times 10^{27}$\\
    &  & 0.1 &V&2.5RA& 4.1 $\times 10^{26}$&-7.8$\times 10^{26}$&8.6$\times 10^{26}$&7.9$\times 10^{24}$&0&4.8$\times 
10^{26}$\\
   &  &0 &V&1.0RA&2.6 $\times 10^{26}$&-4.9$\times 10^{26}$&0&2.1$\times 10^{26}$&0 &-2.0$\times 10^{25}$\\
  &  & -0.1 &V&1.0RA& 2.6$\times 10^{26}$&-1.2$\times 10^{25}$&-2.2$\times 10^{27}$&2.1$\times 10^{26}$&0&-1.7$\times 
10^{27}$\\
  &  & -1.0 &D&5.5RA& 6.1 $\times 10^{26}$&0&-2.3$\times 10^{27}$&-9.4$\times 10^{23}$&0&-1.7$\times 10^{27}$\\\hline
  % % % % % % % % % % % % % % % % % % % %   
 & 1.2&1.0 &V&6.5RA& 8.0 $\times 10^{26}$ &-1.5$\times 10^{27}$&2.1$\times 10^{27}$&-6.7$\times 10^{23}$&2.6$\times 
10^{24}$&1.4$\times 10^{27}$\\
 & &0.1 &V&2.0RA&4.4 $\times 10^{26}$&-8.3$\times 10^{26}$& 1.0$\times 10^{27}$&2.2$\times 10^{25}$&1.5$\times 
10^{25}$&6.5$\times 10^{26}$\\
 & &0&V&1.0RA&3.1 $\times 10^{26}$ &-5.9$\times 10^{26}$&0&2.2$\times 10^{26}$&4.4$\times 10^{25}$&-1.6$\times 
10^{26}$\\
 & &-0.1 &V&1.0RA&3.1 $\times 10^{26}$ &-2.9$\times 10^{25}$&-2.5$\times 10^{27}$&2.2$\times 10^{26}$&6.4$\times 
10^{25}$&-1.9$\times 10^{27}$\\
 & &-1.0 &D&4.7RA&6.8 $\times 10^{26}$ &0&-3.3$\times 10^{27}$&-1.9$\times 10^{23}$&1.6$\times 
10^{25}$&-2.6$\times 10^{27}$\\\hline\hline\label{table}
\end{tabular}
\end{center}
\end{minipage}
\end{table*}
% % % % % % % % % % % % % % %  
From Table \ref{table}, we see that there is a torque reversal for $\xi>1$ and $\xi<1$. 
This results from PGF changing direction whenever the disc transits to and from quasi-Keplerianity regime. 
For $\xi=0.8$, PGF is directed away from the star, resulting into a negative torque which couples with viscous torque. 
This coupled torque is responsible for transporting angular momentum outwards from the neutron star. 
On the other hand, when $\xi=1.2$ the azimuthal velocity is faster and the positive PGF (directed towards the star) 
torque is coupled with the advective torque. In this case angular momentum is advected out of the inner edge of the 
accretion disc, causing the star to spin-up.
Both $N_{adv}(R_{in})$ and $N_{vis}(R_{in})$ can result in warping of the disc \citep{Scott2014}.

Additionally, it is observed that in a quasi-Keplerian system, when $\epsilon\neq0$ the internal dynamo 
generated torque is dominant. 
This is in agreement with the findings of \cite{Tessema2010} for a purely Keplerian disc model with a dynamo.
When $\epsilon=0$, $N_{PGF}$ makes a significant contribution to the total torque. Thus, our mechanism can account for 
the observed enhanced torque reversals in some astronomical environments.

% % % % % % % % % % % % % % % % % % 
\subsection{Comparison with observational results}
Torque reversal from spin-up to spin-down of a neutron star is a common phenomenon. It occurs in systems like 4U 
1626-67 which is observed to have a spin-up/-down rate 
$\dot{\nu}$ as $+8.5\times 10^{-13}$/$-7.0\times 10^{-13}$ Hzs$^{-1}$ at a spin period of 7.6s \citep{Camero2010}. 
Also 4U 1728-247 has a spin-up/-down rate of $\dot{\nu}$ as $+6.0\times 10^{-12}$/$-3.7\times 10^{-12}$ Hzs$^{-1}$ at a 
spin period of 120s  \citep{Bildsten1997}.
These spin variations are related to torque, $N_{\rm{Total}}$ as:
\begin{equation}
 \dot{\nu}=\frac{N_{\rm{Total}}}{2\pi I},
\end{equation}
where $\dot{\nu}$ is the rate of spin change measured in Hzs$^{-1}$, $I$ is the moment of inertia of the neutron star 
defined as 
\begin{equation}
 I=\frac{2}{5}M_s R^2_s.
\end{equation}
The observed spin-up/-down rates are in agreement with the result of our model in Table (\ref{table}) for a neutron star 
with a radius $R_s=10$km \citep{Frank2002}.
% % % % % % % % % 
\section{Conclusion}
We have obtained a complete structure of a quasi-Keplerian model where the magnetic field dynamo forms part. In this 
model, we argue that pressure gradient is not negligible as previous models assumed. Our results show that at 
large radii, the disc remains Keplerian, while inside a critical radius, the rotation is quasi-Keplerian. 
While in this state the accretion disc can make a transition to and from a Keplerian fashion. 
The corotation radius is shifted inwards (outwards) for $\xi>1$ (for $\xi<1$) and the position of the 
inner edge with respect to the new corotation radius is also relocated accordingly as compared to the Keplerian model. 
The resulting torques are of greater magnitude compared to the Keplerian model. 
The interesting part found of a quasi-Keplerian model, is that PGF torque couples with viscous torque 
(when $\xi<1$) to provide a spin-down torque and a spin-up torque (when $\xi>1$) by coupling with the advective torque. 
This enhanced torque reversal is important in explaining the observed variations in 
spin frequency of accretion powered systems like 4U 1626-67.
Further, the dynamo action is in conformity with previous results, except that in a quasi-Keplerian model $N_{dyn}$ 
is of increased magnitude. 
This result is a break through, since finding a complete structure for a quasi-Keplerian disc model has not always been a 
success e.g. \citep{Hoshi1977}.

\begin{acknowledgements}
The authors are grateful to International science programme (ISP) for funding the project and Entoto Observatory and 
Reseach Centre (EORC) for granting a conducive working environment. The authors also acknowledge the comments by the 
referee that have improved the quality of the paper. 
\end{acknowledgements}

\end{document}